\documentclass[a4,aps,twocolumn,showpacs,preprintnumbers,amslatex,amsmath,amssymb]{revtex4-1}
\usepackage{fancyhdr}
\usepackage{epsfig}
\usepackage{graphicx}
\usepackage{verbatim}
\usepackage{indentfirst}
\usepackage{latexsym}
\usepackage{dsfont}

\usepackage{color,soul}

\usepackage{titlesec}
\usepackage{colortbl}

\usepackage[colorlinks,breaklinks]{hyperref}
\usepackage{hyphenat}

\usepackage{natbib}
\DeclareMathOperator{\Tr}{Tr}

\begin{document}

\title{Numerical studies of entanglement properties in one- and two-dimensional quantum Ising and XXZ  models}
\date{\today}
\author{B. Braiorr-Orrs$^1$, M. Weyrauch$^1$, and M. V. Rakov$^2$}
\affiliation{$^1$ Physikalisch-Technische Bundesanstalt, Bundesallee 100, D-38116 Braunschweig, Germany}
\affiliation{$^2$ Kyiv National Taras Shevchenko University, 64/13 Volodymyrska st., Kyiv 01601, Ukraine}

\begin{abstract}
  We investigate entanglement properties of infinite 1D  and 2D  spin-1/2 quantum Ising and XXZ models. Tensor network methods (MPS in 1D and TERG and CTMRG in 2D) are used to model the ground state of the studied models. Different entanglement measures, such as one-site entanglement entropy, one-tangle, concurrence of formation and assistance, negativity and entanglement per bond are calculated and their `characterizing power' to determine quantum phase transitions is compared. A special emphasis is made on the study of entanglement monogamy properties.
\end{abstract}

\pacs{64.70.Tg, 03.67.Mn}

\maketitle

\section{Introduction}

Matter comes in different phases and usually
one can switch between them by changing the temperature.
Close to zero temperature thermal fluctuations disappear and quantum fluctuations dominate.
In this case, by changing a corresponding control parameter one can induce quantum phase transitions (QPTs) between the different ground states of quantum systems. QPTs occur in many different physical systems, and they attract a lot of  attention in condensed-matter physics~\cite{sachdev:book}.

The reason for the recent surge of interest into QPTs are new and exotic quantum phases and critical points, which cannot be described within Landau's theory of phase transitions, i.e. they cannot be characterized by an order parameter. Examples are topologically ordered phases~\cite{wen:book}, quantum spin liquids~\cite{diep:book}, or deconfined quantum critical points~\cite{Senthil05032004, PhysRevB.70.144407}.

At critical points different parts of the system are quantum mechanically strongly correlated and various correlation functions show singular behaviour~\cite{RevModPhys.80.517}.
Several years ago, in quantum information theory it was suggested that quantum phases and QPTs can be characterized and distinguished in terms of quantum entanglement~\cite{PhysRevA.66.032110, Osterloh2002}.

In the present paper we study how different entanglement measures characterize ground state phases
within simple one dimensional and two dimensional spin models. We generate these ground states using
tensor network techniques. Special emphasis is paid to properties of the entanglement monogamy as
expressed through the Coffman-Kundu-Wootters (CKW)~\cite{PhysRevA.61.052306} inequality or similar inequalities.

Bipartite entanglement is the most studied type of entanglement in quantum information theory~\cite{horodecki:2009}, however even bipartite  entanglement measures are under active development, especially with respect to the `identification power' of exotic quantum phases. Recently, the concept of `entanglement spectrum' was introduced~\cite{PhysRevLett.101.010504} and is now studied intensively~\cite{PhysRevB.87.241103, PhysRevLett.110.260403}.

Numerous different measures have been proposed to quantify entanglement~\cite{DBLP:journals/qic/PlenioV07}.
Those which are studied here~\cite{Syljuasen200425, PhysRevA.71.052322, 1367-2630-8-4-061, PhysRevA.81.032304, RevModPhys.80.517} are listed in the Appendix. Studies of the entanglement properties of the many-body systems mainly use the entanglement entropy, the one-tangle, the concurrence, and the fidelity~\cite{RevModPhys.80.517}. There are also investigations of the multipartite entanglement properties of the states (e.g., tripartite entanglement~\cite{dechiara:multient} and global entanglement~\cite{PhysRevA.81.032304}).

It was found in previous studies~\cite{RevModPhys.80.517} that entanglement measures are able to determine critical properties of the systems, in particular the positions of the critical points. Recent studies show that it is also possible to extract the critical exponents, as was e.g. done using finite size scaling of the Schmidt gap~\cite{PhysRevLett.109.237208}. The techniques to extract critical exponents from different entanglement measures are still under development.

In order to simulate the ground states of 1-dimensional and 2-dimensional spin models we use a tensor network (TN) approach
~\cite{1751-8121-42-50-504004, doi:10.1080/14789940801912366}. For a recent review see Ref.~\cite{orus:review}. The basic idea of TN methods is to represent the wave function of a many-body quantum system by a network of interconnected tensors. Experience shows that the TN class of numerical methods is rather flexible~\cite{orus:review}. TNs can handle systems in different dimensions, of finite and infinite size, with different boundary conditions and symmetries. They are able to model systems of bosons, fermions or frustrated spins and can address different types of phase transitions. The  market of TN methods now provides several tens of different methods~\cite{orus:review}, and each of them has its own advantages, disadvantages and areas of applicability.

Matrix product states (MPS) are the most famous among the TN states~\cite{doi:10.1080/14789940801912366}. Powerful algorithms such as  the Density Matrix Renormalization Group (DMRG) ~\cite{PhysRevLett.69.2863} or  Time-Evolving Block Decimation (TEBD)~\cite{vidal:3} can be formulated in terms of MPS. The two-dimensional generalization of the matrix product states is called projected entangled pair states (PEPS)~\cite{Verstraete:arxiv1}. Details about the PEPS and the MPS can be found in Refs.~\cite{PhysRevB.87.085130, PhysRevLett.101.090603, PhysRevB.81.174411, SCH10}.

There is a large variety of TN methods available in order to determine the  quantities which characterize a quantum state (order parameters, critical exponents, entanglement measures). Two questions arise: (1) which numerical method is best suited for the simulation of the ground state of a particular model? (2) which quantity is most efficiently calculated in order to characterize this ground state?

In the present paper we aim to give some input for the answer of these questions using the quantum Ising model in a transverse field and XXZ models in 1D and 2D geometries. We use numerical methods which are able to treat models in the thermodynamic limit: MPS~\cite{SCH10} in 1D and TERG~\cite{gu:4}, CTMRG~\cite{jordan:1, PhysRevB.80.094403} for PEPS in 2D. We compare our results with results from other studies based on other techniques~\cite{Syljuasen200425,PhysRevA.71.052322,1367-2630-8-4-061}. More specifically, we use imaginary-time evolution (the one-site update~\cite{OEST95, OST1997}, TEBD~\cite{VID2007} in 1D and the `simple update' scheme~\cite{PhysRevLett.101.090603} in 2D) to find the approximate ground states for the models under investigation. Exploiting translational invariance of the ground states enables algorithms with
reasonable requirements for computational resources~\cite{levin:1, gu:4}.

As for the second question we calculate a set of different entanglement measures such as one-site entanglement entropy, one-tangle, concurrence of formation and assistance, bounds on localizable entanglement, local entanglement, negativity and entanglement per bond and compare their `characterization power' of the state of the system.
Moreover, we will compare the entanglement properties of the chosen models in 1D and 2D geometries.
In this way we obtain information on the `monogamy of entanglement' or entanglement distribution~\cite{PhysRevA.61.052306}. The entanglement monogamy properties will be studied in some detail.

This paper is organized as follows.  In Sec.~\ref{numer_methods}, the one-site and two-site reduced density matrices (needed for the entanglement measures calculation) are determined from the MPS and PEPS representations of the ground states. Sec.~\ref{results} presents numerical results and their interpretations for quantum Ising and XXZ models.  Conclusions are made in Sec.~\ref{conclu}. Various entanglement measures are briefly listed and discussed in the Appendix.

\section{Translationally invariant tensor network methods}\label{numer_methods}

In this section we briefly describe the different tensor network methods we
use to obtain the translationally invariant ground state wave functions for various spin models.
We describe the renormalization steps one needs to take in order to prevent exponential
increase of the bond dimensions of the tensors for 2D systems. Furthermore,
we show how reduced density matrices are determined using the tensor entanglement renormalization
group (TERG) or the corner transfer matrix renormalization group (CTMRG).
We also briefly discuss a renormalization technique for translationally invariant 1D systems for comparison.
All these methods are closely related conceptually,
but differ in many technical details.

\subsection{Imaginary-time evolution}\label{sec:ite}

Imaginary-time evolution is one method for finding ground-state wave functions~\cite{CPA:CPA3160070404}.
It evolves an arbitrary state  $|\psi\rangle$  (which contains the ground state as a component)
into the ground state $|\psi_0 \rangle$ of the Hamiltonian $H$,
\begin{equation}
|\psi_0 \rangle=\lim_{\tau \rightarrow \infty}\frac{\exp(-\tau H)|\psi\rangle}{\|\exp(-\tau H)|\psi\rangle\|}.
\end{equation}
Strictly speaking, this is correct only if the ground state is non-degenerate. If the ground state is degenerate one obtains an (arbitrary) linear combination of the degenerate ground states. Since the imaginary time evolution operator is not unitary, normalization must be explicitly ensured through the denominator in the above expression.

\begin{figure}[h!]
\unitlength 1cm
    \includegraphics[width=0.25\textwidth]{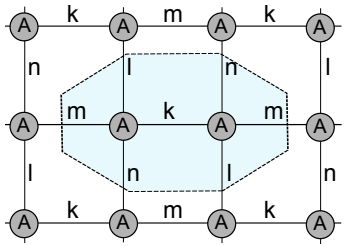}
    \caption{\footnotesize Graphical representation of a translationally invariant PEPS. All tensors $A$ are equal. The letters denote the different bond types.  For a finite lattice the open ends are connected: left with right and top with bottom.  The octagon encloses the tensors $M$ or $\bar{M}$ defined in Eqs.~(\ref{Mtensor}) and (\ref{Mbartensor}), respectively.}
    \label{fig:directionsInv}
\end{figure}

We will first describe imaginary-time evolution of the translation invariant PEPS in two dimensions.
In both dimensions we use periodic boundary conditions.
The initial  random wave function $|\psi\rangle$ is constructed from a product of equal rank-5 tensors $A$ at the lattice sites $i,j,\ldots$,
\begin{equation}\label{wave-function-2}
|\Psi\rangle = {\rm tTr} \left\{ A^{\sigma_i}_{k_il_im_in_i} A^{\sigma_j}_{k_jl_jm_jn_j}\ldots |\sigma_i \sigma_j \ldots \rangle \right\}.
\end{equation}
The tensors contain random entries. The size of each physical (spin) index $\sigma$ is two, since we consider spin-1/2 systems only. The size of each virtual bond $k, l, m, n$ is $D$. The tensor trace ${\rm tTr}$ includes  summation over all spin configurations and over all bond indices. The tensor network describing
this state is graphically represented in Fig.~\ref{fig:directionsInv}.

We consider spin systems described by Hamiltonians with nearest neighbor interactions only.
Therefore, the Hamiltonian may be decomposed into four terms,
\begin{equation}
H=H_{k}+H_{l}+H_{m}+H_{n}.
\end{equation}
These terms correspond to the four bond types shown in Fig.~\ref{fig:directionsInv}.
Single-site operators in the Hamiltonian may  be easily incorporated into these terms.
The nearest neighbor interactions within each of the four terms commute with each other. However, the four terms $H_k$, $H_m$, $H_l$ and $H_n$ do not commute. As a consequence, we cannot write the evolution operator $P_\tau=\exp(-\tau H)$ as a product of two-site operators.

However, we may use a first-order Trotter-Suzuki expansion~\cite{trotter}
\begin{equation}
e^{-\Delta\tau H}\approx e^{-\Delta\tau H_{k}}  e^{-\Delta\tau H_{l}}  e^{-\Delta\tau H_{m}}  e^{-\Delta\tau H_{n}} + O(\Delta\tau^2).
\end{equation}
with  a `small' imaginary time step $\Delta\tau$ in order to write $P_\tau$ approximately as a product of two-site operators
\begin{equation}
P_\tau \approx \prod_{\substack{\scalebox{0.7}{time steps}  \\ \scalebox{0.7}{ sites } }} e^{-\Delta\tau h^{ij}_{k}}  e^{-\Delta\tau h^{ij}_{l}}  e^{-\Delta\tau h^{ij}_{m}}  e^{-\Delta\tau h^{ij}_{n}}.
\end{equation}
where the $h_\lambda^{ij}$ represent the nearest neighbor interactions of type $\lambda$ between site $i$ and $j$
($H_\lambda=\sum_{ij}h_\lambda^{ij}$).
Since  imaginary-time evolution is a projective method, Trotter errors do not accumulate
during the evolution~\cite{bamler:1}.
Higher orders of the Trotter-Suzuki expansion may be used in order to achieve a better of convergence.

In practice, we repeatedly apply imaginary time step operators for different `directions' $k$, $l$, $m$, or $n$  to a random state until convergence is achieved. Convergence is judged using suitable criteria, $\Delta\tau$ and the total number of time steps $N$ are chosen in order to achieve the desired accuracy. We start from $\Delta\tau=0.1$ as the initial time step and then progressively reduce it until the state does not change any more within specified limits. We take this converged state as our approximation for the ground state $|\psi_0\rangle$.

\subsection{Update schemes}

After each imaginary time step the size of the tensors describing the state increases. Let us consider the evolution of the state over a $k$ bond. Assuming that the tensors $A^{\sigma_i}_{k_il_im_in_i}$ and $A^{\sigma_j}_{k_jl_jm_jn_j}$ correspond to neighboring sites connected by a $k$-bond, the two-site tensor $M$ becomes
\begin{equation}\label{Mtensor}
 M^{\sigma_i\sigma_j}_{ l_im_in_i l_jm_jn_j}=\sum^{D}_{k=1}  A^{\sigma_i}_{kl_im_in_i}A^{\sigma_j}_{kl_jm_jn_j}.
\end{equation}
Applying the two-site operator $p_k=e^{-\Delta\tau h^{ij}_{k}}$ to $M$ produces the `evolved' two-site tensor
\begin{equation}\label{Mbartensor}
 \bar{M}^{\sigma_i'\sigma_j'}_{l_im_in_i l_jm_jn_j}=\sum^{d}_{\sigma_i \sigma_j} (p_k)^{\sigma_i'\sigma_j'}_{\sigma_i\sigma_j} M^{\sigma_i\sigma_j}_{ l_im_in_i l_jm_jn_j}
\end{equation}
of the same size as $M$.
However, reconstructing from this tensor the PEPS tensors $A$ using a singular value decomposition
\begin{equation}
\bar{M}_{(\sigma_i'l_im_in_i)(\sigma_j' l_jm_jn_j)}=\sum^{d\times D^3}_{\tilde{k}=1} U_{(\sigma_i'l_im_in_i)\tilde{k}}\Lambda_{\tilde{k},\tilde{k}}V^{\dagger}_{\tilde{k}(\sigma_j' l_jm_jn_j)}
\label{growth}
\end{equation}
increases the size of the $k$-bond from $D$ to $d\times D^3$.

In order to prevent an exponential growth of the tensor size during the imaginary-time evolution,
the bond size must be suitably reduced at each time step. This should be done in such a way that the difference between the evolved state $|\psi^{\text{evol.}}\rangle$ with increased dimension and its approximation with reduced dimension $|\psi^\text{approx.}\rangle$, $K=\||\psi^\text{evol.}\rangle-|\psi^\text{approx.}\rangle \|$, is minimal.
An imaginary time step together with the necessary reduction of the tensor size is called an `update step', and the corresponding method to reduce the bond size is called `update scheme'.

Different types of update schemes exist. In general, in order to implement an update step one needs to take into account the whole environment of two evolving tensors~\cite{orus:review}. Update schemes that act this way are called `full updates'. They are numerically rather costly.  Less demanding on the computational resources is the `simple update' scheme~\cite{PhysRevLett.101.090603}, which takes the environment into account approximately; it is based on a generalization of a method developed for 1D systems called `time-evolving block decimation' (TEBD)~\cite{vidal:3}.  There is also a cluster update scheme~\cite{PhysRevB.86.195137}, which compromises between the two previously discussed schemes.

In the present work we use the `simple update' scheme following Refs.~\cite{PhysRevLett.101.090603,bamler:1}.
Here, we would like to comment on two important aspects of this algorithm.

\begin{figure}[h!]
\unitlength 1cm
    \includegraphics[width=0.3\textwidth]{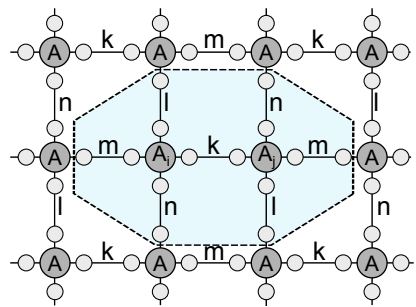}
    \caption{\footnotesize Graphical representation of a translationally invariant PEPS with bond vectors. Small circles correspond to $\sqrt{\lambda_{\alpha}}$, $\alpha=k,l,{m},{n}$. The octagon defines the object to be renormalized. The extra $\sqrt{\lambda_\alpha }$ taken from the external tensors are used to mimic renormalization effects due to the environment.}
    \label{fig:simpleUpdate}
\end{figure}

In the `simple update' scheme one introduces bond vectors $\lambda_k$, $\lambda_l$, $\lambda_m$, $\lambda_n$ in addition to the tensors $A$ of the standard PEPS. The connection between the tensors introduced in~\eqref{wave-function-2} (now we denote these tensors by $\bar{A}^{\sigma}_{klmn}$) and the tensors of new representation  $A^{\sigma}_{klmn}$ is given by $\bar{A}^{\sigma}_{klmn}=A^{\sigma}_{klmn}\sqrt{\lambda_k}\sqrt{\lambda_l}\sqrt{\lambda_m}\sqrt{\lambda_n}$.
A graphical representation of the translationally invariant PEPS with additional bond vectors is shown in Fig.~\ref{fig:simpleUpdate}. The small circles in the figure correspond to $\sqrt{\lambda_{\alpha}}$, $\alpha={k},{l},{m},{n}$. The new PEPS representation directly corresponds to the canonical form of MPS introduced by Vidal~\cite{vidal:3}.

While the introduction of bond vectors appears to be a trivial redefinition of the local tensors, it is their role
in the renormalization or update scheme, which will prove to be non-trivial: naively, one would assume that it is
$\bar{M}$ defined in Eq.~(\ref{Mbartensor}) which is renormalized such that the size of the PEPS tensors do not grow. However, it is the tensor $S$ defined by
\begin{equation}
\begin{split}
S^{\sigma_i' \sigma_j'}_{l_im_in_i l_im_in_i}&= \sqrt{\lambda_{l_i}}\sqrt{\lambda_{m_i}}\sqrt{\lambda_{n_i}}  \times \\
\bar{M}^{\sigma_i'\sigma_j'}_{l_im_in_i l_jm_jn_j} &
   \sqrt{\lambda_{l_j}}\sqrt{\lambda_{m_j}}\sqrt{\lambda_{n_j}}.
\end{split}
\end{equation}
which is renormalized.  Here, $\bar{M}$ is defined as in Eq.~(\ref{Mbartensor}) but $M$ includes all factors $\sqrt{\lambda}$ necessary due to the redefinition of the $A$ tensors,
\begin{equation}
\begin{split}
 M^{\sigma_i\sigma_j}_{ l_im_in_i l_jm_jn_j}= & \sqrt{\lambda_{l_i}}\sqrt{\lambda_{m_i}}\sqrt{\lambda_{n_i}}\times \\ \sum^{D}_{k=1}& A^{\sigma_i}_{kl_im_in_i}\lambda_k A^{\sigma_j}_{kl_jm_jn_j}  \sqrt{\lambda_{l_j}}\sqrt{\lambda_{m_j}}\sqrt{\lambda_{n_j}}.\nonumber
 \end{split}
\end{equation}
It is important to note that $S$ contains extra factors of $\sqrt{\lambda}$ taken from the environment
as indicated by the octagon in the graphical representation of the tensor $S$  in Fig.~\ref{fig:simpleUpdate}.
Such an approach is called  `mean field approximation' of the environment. However, this statement is intuitive and lacks mathematical rigour. The procedure is justified by numerical success. Vidal has proved a number of statements which justify this procedure in 1D for TEBD~\cite{vidal:3}.

There is another aspect of the simple update scheme that requires comment. Originally the `simple update' method was implemented for studying quantum models on a honeycomb lattice~\cite{PhysRevLett.101.090603}, where bipartition of the lattice is necessary, i.e. the ground state in PEPS form is given by two tensors $A$, $B$ representing two sublattices.

One would expect that on a square lattice with translational invariance the ground state must be translationally invariant too, i.e. no bipartition is expected. However, in various papers~\cite{jordan:1, PhysRevB.80.094403,PhysRevB.87.085130} which use `simple update' to simulate the ground state on a square lattice, a bipartition is introduced even for translationally invariant models. It is stated in Ref.~\cite{PhysRevB.87.085130} that imaginary-time evolution breaks translational invariance of the lattice. The motivation of the bipartition is not explained clearly in the literature, thus we want to highlight this aspect of the `simple update' scheme.

Our experience shows that, if we start our evolution with a random translationally invariant state given by random tensor $C$ sitting on each site, the `simple update' leads to a translationally invariant ground state, too. The equality of $A$ and $B$ tensors at the end of the imaginary-time evolution is very dependent on the convergence criteria and on the bond size $D$ of the PEPS tensors.
The resulting $A$ tensors in this case are rotationally symmetric with high accuracy (rotation corresponds to cyclic permutation of virtual bond indices). Rotational symmetry is also ensured by the approximate equality of four bond vectors. The accuracy of their equality is given by the convergence condition. Note, that for higher $D$ it is much harder numerically to obtain approximately equal $A$ and $B$.

We suggest that the resulting translational invariance could be highly dependent on the numerical implementation of the singular value decomposition procedure
used during imaginary-time update. In practise, due to a gauge freedom the simple update scheme can lead to a bipartitioning of the lattice in general,
i.e. translational invariance would be superficially  broken. In fact, translational invariance is maintained and could be restored explicitly using an appropriate transformation.

The gauge freedom  can be easily demonstrated for a product of two equal matrices,
\begin{equation}
M=A A = (A \Lambda) (\Lambda^{-1}A) = C D.
\end{equation}
The SVD which is used within the simple update scheme as indicated in Eq.~\eqref{growth}
\begin{equation}
M=U\Lambda V^{\dag}=(U\sqrt{\Lambda} )( \sqrt{\Lambda}V^\dag)= C D.
\end{equation}
leads to the purely numerical bipartitioning of the tensors on the lattice.

Moreover, we found that the `simple update' can distinguish ferromagnetic and antiferromagnetic order in the state. This order is defined by the sign of the coupling constant in the two-site Hamiltonian that is used during the evolution. Thus, the usage of antiferromagnetic coupling constant will lead to $A$ and $B$ tensors that differ with respect to spin-flip transformation.

Therefore, as an output of the simple update scheme for translationally invariant models we obtain the PEPS given by {\it one rank-5 tensor} $A^{\sigma}_{klmn}$ and {\it just one unique bond vector} $\lambda=\lambda_k=\lambda_l=\lambda_m=\lambda_n$. After the completion of the imaginary time evolution, we multiply the bond vectors into the tensors, $\bar{A}^{\sigma}_{klmn}=A^{\sigma}_{klmn}\sqrt{\lambda_k}\sqrt{\lambda_l}\sqrt{\lambda_m}\sqrt{\lambda_n}$, since the bond vectors are not needed any more. The resulting tensor network which will be used for further tensor contraction algorithms is the same as shown in the Fig.~\ref{fig:directionsInv}, however, with new $A$ tensors sitting on each site.

\begin{figure}[h!]
    \centering
    \includegraphics[width=0.25\textwidth]{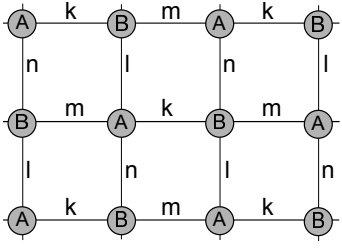}
    \caption{\footnotesize After the imaginary-time evolution in the case of not purely translationally invariant state (i.e. describing antiferromagnetic state) the tensor network assumes
              a bipartitioned structure in terms of tensors $A$ and $B$.}
    \label{fig:bipartition}
\end{figure}

If the ground state is not purely translationally invariant, i.e. has the antiferromagnetic order, PEPS representation would then require two tensors $A$ and $B$ to describe the state. Here, after the completion of the imaginary time evolution, we multiply the bond vectors into the tensors $A$ and $B$ again. Thus, in the case of lattice bipartition the sublattices corresponding to these tensors are denoted as $\mathcal{A}$ and $\mathcal{B}$, respectively. The resulting tensor network which will be used from now on is shown in Fig.~\ref{fig:bipartition}.

In the following we will denote tensors obtained from imaginary-time evolution and with incorporated bond vectors $\lambda_{\alpha}$ by $A$ and $B$ without bars for simplicity.

\subsection{Reduced density matrices for 2D systems}

In this subsection we briefly present two different methods for the calculation of the $n$-spin reduced density matrices $\rho_n$ for 2D systems ($n \leq 4$). From the reduced density matrices we obtain the expectation value of an $n$-spin operator $O_n$  in the standard way: $\langle \hat{O_n} \rangle = {\rm Tr}(\hat{O_n}\rho_n)$. The calculation of the density matrices in a tensor network approach involves a tensor trace, the calculation of which is exponentially hard in 2D and, therefore, requires renormalization methods (in contrast, for 1D systems the calculation of the reduced density matrices can be achieved  in polynomial time). The methods we discuss here are the tensor-entanglement renormalization group (TERG)~\cite{gu:4} and the corner transfer matrix renormalization group (CTMRG)~\cite{jordan:1, PhysRevB.80.094403}.

The essentials of these methods are described e.g. in the papers cited above for the calculation of expectation values. Here, we present these methods for the calculation of reduced density matrices.

\subsubsection{Tensor-entanglement renormalization group}

\begin{figure}[h!]
    \includegraphics[width=0.3\textwidth]{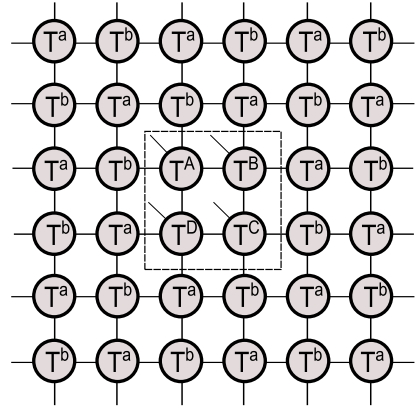}
    \vskip 3mm
    \caption{\footnotesize Tensor network and impurity tensors used in TERG. The tensor network is in principle infinitely large. The open lines at the impurity sites $A, B, C, D$ in the boxed center of the figure indicate the physical spin indices of the impurity tensors. The open lines at the boundary of the figure are connected to tensors not shown.}
    \label{fig:terg_lattice}
    \vskip 0.4cm
\end{figure}

TERG is based on the tensor renormalization group (TRG) method introduced by Levin and Nave~\cite{levin:1} for classical systems. It was modified for quantum systems in Ref.~\cite{gu:4} using the concept of `impurity' tensors. In the present paper, we name `impurity' positions in a tensor network  those positions at which spin operators are attached or
where the physical indices of the network are explicitly kept. At all other positions the physical indices are summed over.
At each site, which is not an impurity site, we define the following tensors (see Fig.~\ref{fig:terg_lattice})
\begin{equation}\label{double layer}
\begin{split}
& T^a_{\bar{k}\bar{l}\bar{m}\bar{n}}=\sum_{\sigma} A^{\sigma *}_{k'l'm'n'}A^{\sigma}_{klmn}, \\
& T^b_{\bar{m}\bar{n}\bar{k}\bar{l}}=\sum_{\sigma} B^{\sigma *}_{k'l'm'n'}B^{\sigma}_{klmn}.
\end{split}
\end{equation}
with the virtual bonds $\bar{k}=k'k$, $\bar{l}=l'l$, $\bar{m}=m'm$, $\bar{n}=n'n$. Each index has dimension $D^2$.

Furthermore, at the impurity sites we define four `impurity' tensors $T^A$, $T^B$, $T^C$, and $T^D$ with  physical bond $\bar{\sigma}=\sigma'\sigma$ of dimension $d^2$ as illustrated in Fig.~\ref{fig:terg_lattice},
\begin{equation}
\begin{split}
 (T^{A})^{\bar{\sigma}}_{\bar{k}\bar{l}\bar{m}\bar{n}}&= (A)^{\sigma' *}_{k'l'm'n'}(A)^{\sigma}_{klmn}, \\
 (T^{B})^{\bar{\sigma}}_{\bar{m}\bar{n}\bar{k}\bar{l}}&= (B)^{\sigma' *}_{m'n'k'l'}(B)^{\sigma}_{mnkl}, \\
 (T^{C})^{\bar{\sigma}}_{\bar{k}\bar{l}\bar{m}\bar{n}}&= (A)^{\sigma' *}_{k'l'm'n'}(A)^{\sigma}_{klmn}, \\
 (T^{D})^{\bar{\sigma}}_{\bar{m}\bar{n}\bar{k}\bar{l}}&= (B)^{\sigma' *}_{m'n'k'l'}(B)^{\sigma}_{mnkl}.
\end{split}
\label{fourABCD}
\end{equation}
The tensors $T^A$ and $T^C$ are located at sites of sublattice $\mathcal{A}$  and tensors $T^B$ and $T^D$ at sites of sublattice $\mathcal{B}$.
For simplicity, from now on we will omit the overbars for the indices labeling the various $T$ tensors and just keep
in mind that the virtual indices have dimensions $D^2$ and the physical index has the dimension $d^2$.

\begin{figure}[h!]
    \includegraphics[width=0.2\textwidth]{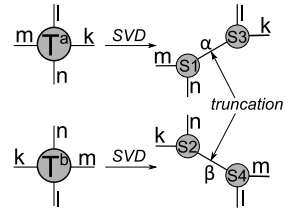}
    \vskip 3mm
    \caption{\footnotesize SVD of the tensors $T^a$ and $T^b$. The indices $\alpha$ and $\beta$ will be
    truncated in order to prevent exponential growth of the tensors $\tilde{T}$ defined in Eq.~(\ref{coarse1})}
    \label{fig:SVD}
\end{figure}

Now, depending on the sublattice, we perform one of the following singular value decompositions (the arrow indicates a reshaping of indices),
\begin{equation}\label{TtoS}
\begin{split}
T^a_{klmn} \rightarrow M^a_{(kl)(mn)}\rightarrow\sum_{\alpha=1}^{D_{c}} S^3_{kl\alpha} S^1_{mn\alpha}, \\
T^b_{mnkl} \rightarrow M^b_{(nk)(lm)}\rightarrow\sum_{\beta=1}^{D_{c}} S^2_{nk\beta} S^4_{lm \beta}.
\end{split}
\end{equation}
The $S^i$ tensors  are obtained from the $U$ and $V^{\dagger}$ tensors of the SVD $M=\sum U \Lambda V^{\dagger}$ by multiplication with $\sqrt{\Lambda}$. These decompositions are illustrated in the Fig.~\ref{fig:SVD}.
Analogous SVDs  are performed for tensors $T^A$, $T^B$, $T^C$, $T^D$, e.g.
\begin{equation}\label{TtoSIm}
\begin{split}
& (T^{A})^{(\sigma_A' \sigma_A)}_{klmn}\rightarrow M^A_{(\sigma_A'kl)(\sigma_A mn)} \rightarrow \sum_{\alpha=1}^{D_{c}} S^{A3}_{\sigma_A'kl\alpha} S^{A1}_{\sigma_A mn\alpha}, \\
& (T^{B})^{(\sigma_B' \sigma_B)}_{mnkl}\rightarrow M^B_{(\sigma_B nk)(\sigma_B' lm)} \rightarrow \sum_{\beta=1}^{D_{c}} S^{B2}_{\sigma_Bnk\beta} S^{B4}_{\sigma_B' lm \beta}.
\end{split}
\end{equation}

\begin{figure}[h!]
    \includegraphics[width=0.35\textwidth]{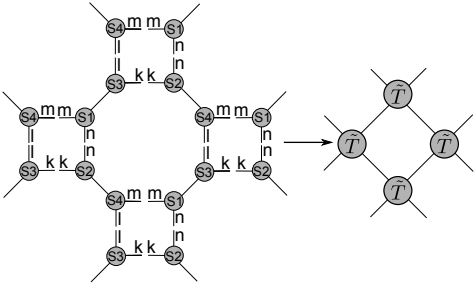}
    \vskip3mm
    \caption{ \footnotesize Coarse graining: Contraction of four $S$ tensors. The number of $T$ tensors
              is reduced by a factor of 2.}
    \label{fig:contractionABCD}
 \end{figure}

The last step of the TERG procedure is coarse-graining, that is the contraction of four $S$ tensors into one $\tilde{T}$  tensor
\begin{equation}\label{coarse1}
\tilde{T}_{\alpha \beta \gamma \delta}=\sum_{klmn} S^{2}_{nk\alpha} S^{1}_{mn\beta} S^{4}_{lm\gamma} S^{3}_{kl\delta}.
\end{equation}
as illustrated in Fig.~\ref{fig:contractionABCD}.

Renormalized $\tilde{T}^A$, $\tilde{T}^B$, $\tilde{T}^C$, $\tilde{T}^D$ tensors are determined in the same way,  e.g.
\begin{equation}\label{coarse2}
(\tilde{T}^{A})^{(\sigma_A' \sigma_B)}_{\alpha \beta \gamma \delta}=\sum_{klmn} S^{B2}_{\sigma_B nk\alpha} S^{1}_{mn\beta} S^{4}_{lm\gamma} S^{A3}_{\sigma_A' kl\delta}.
\end{equation}
using the decompositions~(\ref{TtoS}) and (\ref{TtoSIm}).

From the Eqs.~(\ref{TtoS}) and (\ref{TtoSIm}) we realize that the size of the virtual bonds $\alpha$ and $\beta$ is $D^4$ and $D^4d^2$, respectively, so that the virtual bonds of the tensors $\tilde{T}$ would grow exponentially without suitable truncation.
In order to prevent the exponential growth we truncate these indices to $D_c$, that is
we neglect small singular values in the expansion Eq.~(\ref{TtoS}) and~(\ref{TtoSIm}); $D_c$ has to be chosen large enough to
maintain the relevant physical information but small enough to stay within acceptable numerical cost. An acceptable choice for $D_{c}$  depends on the virtual dimension $D$ of the PEPS. In our calculations we use $D_{c}$ between $16$ and $64$.

After a sufficient number of the TERG transformations as described above the tensors $\tilde{T}^A$, $\tilde{T}^B$, $\tilde{T}^C$, $\tilde{T}^D$ contain all relevant information necessary to calculate observables, e.g.
the four-site reduced density matrix
\begin{equation}
\rho_{\sigma_A \sigma_B \sigma_C \sigma_D \sigma_A' \sigma_B' \sigma_C' \sigma_D'} =
{\rm Tr} (\tilde{T}^A \tilde{T}^B \tilde{T}^C \tilde{T}^D),
\label{reduced4}
\end{equation}
which has to be normalized such that ${\rm Tr} \rho=1$. The trace includes summations over virtual indices. Two-site and one-site reduced density matrices are then easily obtained from $\rho$ by a partial trace.

TERG transformations are applied until a convergence condition is satisfied. After each TERG transformation we calculate $\rho$ until we find $\|\rho^{(n)}-\rho^{(n+1)} \|<\varepsilon$, where $\rho^{(n)}$ denotes the reduced density matrix $\rho$ at the $n$-th recursion step. The matrix norm is implemented as $\|X\|=\sqrt{\sum_{ i,j}x_{ij} ^2}$.  In practice, we take  $\varepsilon$ between $10^{-5}$ and  $10^{-10}$.

From the two-site and single-site reduced density matrices we calculate the desired physical quantities in section~\ref{results}.

\subsubsection{Corner transfer matrix renormalization group}

\begin{figure}[h!]

    \includegraphics[width=0.45\textwidth]{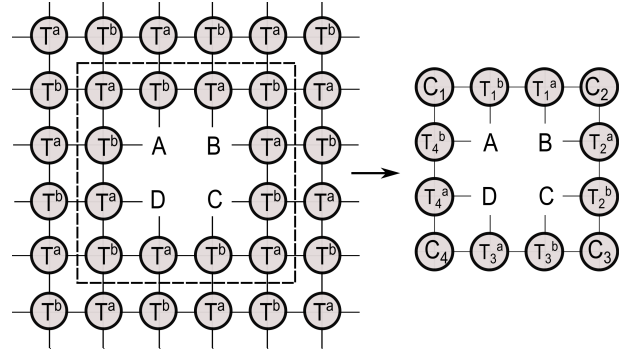}
    \vskip 3mm
    \caption{\footnotesize Tensor network used in CTMRG (left) and environment tensor (right).
             Here, there are no tensors at the sites $A, B, C, D$.}
    \label{fig:environment4}
\end{figure}

The corner transfer matrix renormalization group (CTMRG) was first introduced by Baxter~\cite{baxter01}. It was further developed and applied to  classical statistical systems by  Nishino and Okunishi~\cite{doi:10.1143/JPSJ.65.891,doi:10.1143/JPSJ.66.3040}. More recently, it was  adapted to the contraction of tensor networks by Orus~\cite{PhysRevB.80.094403,PhysRevB.85.205117}. CTMRG determines the `environment tensor'
$\mathcal{G}$ of the four sites $A, B, C, D$ as defined in Fig.~\ref{fig:environment4}. The locations of these four sites correspond to the locations of the `impurity sites' in TERG. The relation between the environment tensor and the four-spin reduced density matrix will be given below.

Similarly to the TERG, one starts from  $T^a$ and $T^b$ (see Eq.~(\ref{double layer})) located at the corresponding sites of the tensor network with the exception of the four sites $A, B, C, D$ (see Fig.~\ref{fig:environment4}).  After the complete contraction of
this tensor network, one obtains twelve tensors $C_1$, $T^b_{1}$, $T^a_{1}$, $C_2$, $T^a_{2}$, $T^b_{2}$, $C_3$, $T^b_{3}$, $T^a_{3}$, $C_4$, $T^a_{4}$, $T^b_{4}$ shown on the right side of Fig.~\ref{fig:environment4}. They constitute the environment tensor. In order to determine them, the CTMRG algorithm successively contracts more and more tensors from the network into these twelve tensors (see Fig.~\ref{fig:environment4}).

\begin{figure}[h!]
    \includegraphics[width=0.4\textwidth]{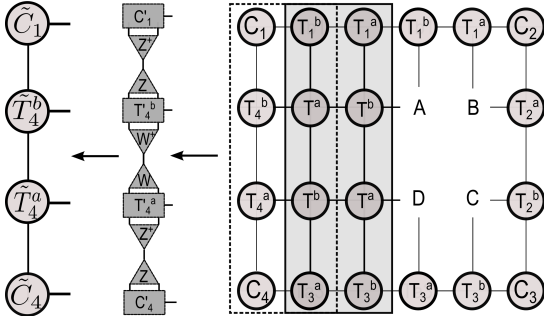}
    \vskip 3mm
    \caption{\footnotesize
    Left move of a CTMRG transformation. The tensors in the shaded box (right part
    of the figure) are inserted into the environment tensor (step~1). The inserted tensors are then combined with
    the left column of tensors (absorption, step~2)
    and  renormalized (step~3) as illustrated in the center figure.
    The renormalized tensors shown in the left part of the figure replace the boxed part of the right part of the figure to form the renormalized environment tensor.}
    \label{fig:left_step}
\end{figure}

In order to prevent exponential growth of the virtual bond size a renormalization
is performed at each step just like in TERG. However, the details of these renormalization steps are somewhat different.
In CTMRG the twelve tensors are renormalized  by left, up, right and down `moves' defined and described
in the following four steps. We describe left moves only as illustrated in Fig.~\ref{fig:left_step}, the others are done analogously. The description of the steps follows Orus~\cite{PhysRevB.80.094403}:

\textit{Step 1.} \textit{Insertion:} insert two sets (columns) of tensors as shown in Fig.~\ref{fig:left_step}. (The insertion of two sets is only necessary because of translational symmetry breaking discussed in the previous subsections.)

\textit{Step 2.} \textit{Absorption:} absorb the first set of tensors into new tensors with increased
vertical bond size: $C'_1=C_1 T^b_{1}$, $T^{b\prime}_{4}=T^b_{4}T^a$, $T^{a\prime}_{4}=T^a_{4}T^b$, $C'_4=C_4 T^a_{3}$.  (Here and in the following we omit the indices of the tensors, since they are easily reconstructed from the corresponding figures.)

\textit{Step 3.} \textit{Renormalization}: Insert two types of approximate isometries $Z$ ($Z^{\dagger}Z\approx I$) and $W$ ($W^{\dagger}W\approx I$) as shown on the left side
in Fig.~\ref{fig:left_step} such that the vertical bond size of the tensors $C'_1, T^{b\prime}_{4}, T^{a\prime}_{4}, C'_4$ is truncated. The renormalized tensors are $\tilde{C}_1=C'_1Z^{\dagger}$, $\tilde{T}^{b}_{4}=Z T^{b\prime}_{4} W^{\dagger}$, $\tilde{T}^{a}_{4}=W T^{a\prime}_{4}Z^{\dagger}$,  $\tilde{C}_4=ZC^\prime_4$, and $I$ is the identity matrix.
One determines $Z$ from an eigenvalue decomposition of the matrix $C_1^\prime C_1^{\prime\dagger} + C_4^{\prime\dagger} C^\prime_4 = \tilde{Z} D_{Z} \tilde{Z}^{\dagger}$ and  $W$  from an eigenvalue decomposition of the matrix $Q^\prime_1 Q_1^{\prime\dagger}+Q_4^{'\dagger}Q_4'=\tilde{W}D_{W}\tilde{W}^{\dagger}$ with $Q_1^\prime=C_1^\prime T^{b\prime}_{4}$, $Q_4^\prime=C_4^\prime T^{a\prime}_{4}$. In order to achieve the desired truncation ($\tilde{Z}\rightarrow Z$, $\tilde{W}\rightarrow W$) one keeps only the eigenvectors belonging to the $D_{c}$ largest eigenvalues of $D_{Z}$  and $D_W$, respectively.

\textit{Step 4.} Repeat steps 2 and 3 for the second set of inserted tensors.

After the absorption and renormalization of the second set of tensors, one obtains the renormalized  tensors $\tilde{C}_1,\tilde{T}^{b}_{4},\tilde{T}^{a}_{4},\tilde{C}_4$ for the left column of tensors of the environment. A sequence of one left, down, right, and up moves constitutes one CTMRG transformation. Obviously, this transformation resembles a coarse-graining of the tensor network. The moves described above are repeated until convergence is achieved. We use the same convergence condition as for TERG with
four-spin reduced density matrix given by
\begin{equation}\label{rhoCTMRG}
\rho_{\sigma_A \sigma_B \sigma_C \sigma_D \sigma_A' \sigma_B' \sigma_C' \sigma_D'} ={\rm Tr}( \mathcal{G} T^A T^B T^C T^D),
\end{equation}
in terms of the environment tensor $\mathcal{G}$ and the {\it unrenormalized} tensors $T^A$, $T^B$, $T^B$, $T^C$
defined in Eq.~(\ref{fourABCD}). Of course, the reduced density matrix has to be normalized such that ${\rm Tr} \rho=1$.
Alternatively, one may renormalize until for some $n$: $\sum_{i=1}^4||\Lambda_i^{(n)}-\Lambda_i^{(n+1)} ||<\varepsilon$, where $\Lambda_i$ is the singular matrix of the corresponding corner tensor $C_i$~\cite{privatOrus}.

In order to start up the recursive renormalization described above, all 12 tensors constituting the environment
are set to tensors $T^a$ and $T^b$, respectively, and superfluous indices are traced out.

A comparison of the two results Eq.~(\ref{reduced4}) and~(\ref{rhoCTMRG}) for the four-spin reduced density matrix may be instructive. In TERG the renormalized `impurity tensors' $\tilde{T}^i$ contain all information about the tensor network, while in CTMRG one finally has to contract the unrenormalized impurity tensors into the renormalized `environment tensor' $\mathcal{G}$ in order to get the density matrix. The renormalization procedures used in both methods in order to prevent exponential growth of indices are somewhat different, however, it becomes obvious from the above descriptions that the two methods are in fact closely related.

\subsection{Translationally invariant matrix product states}

In this section we will briefly describe the methods we use for 1D systems. Calculations in 1D are numerically far less demanding than 2D calculations. However, it is instructive to compare different methods.

The most efficient methods for 1D calculations are variational methods. They are described in detail by Schollw{\"o}ck~\cite{SCH10}. Various imaginary time evolution algorithms have also been investigated for 1D,
notably the TEBD algorithm proposed by Vidal~\cite{VID2007}. This algorithm motivated the 2D algorithm described in section~\ref{sec:ite}.
As discussed there, the TEBD method locally breaks translational invariance.

Here, however, we would like to discuss a method which maintains translational invariance exactly, i.e. we represent a state $|\Psi\rangle$ of $N$ spins by a matrix product state (MPS) with identical matrices $A^\sigma$ at each lattice site
\begin{equation}
|\Psi\rangle=\sum_{\sigma_1,\ldots,\sigma_N} {\rm Tr} (A^{\sigma_1}\cdot A^{\sigma_2}\cdot
\ldots \cdot A^{\sigma_N} )|\sigma_1,\ldots,\sigma_N\rangle.
\end{equation}
The rank-3 tensors $(A^\sigma_{p,p^\prime})$ have physical (spin) index $\sigma$ of size two (since we consider spin-1/2 systems only) and virtual dimensions of size $m$.
Such an MPS was already introduced in the seminal papers by {\"O}stlund and Rommer~\cite{OEST95, OST1997}; the PEPS
introduced in Eq.~(\ref{wave-function-2}) is its straightforward 2D generalization. We assume periodic boundary conditions.

In order to implement imaginary time evolution without locally breaking translation invariance one requires a matrix product operator (MPO) representation of the time evolution operator $\exp(-\tau H)$,
\begin{eqnarray}
\exp(-\tau H) = &&\sum_{\mbox{\scriptsize\parbox[c]{1.5cm}{$\sigma_1,\ldots,\sigma_N,$\\
$\sigma^\prime_1,\ldots,\sigma^\prime_N$}}}
{\rm Tr} \left( W^{\sigma_1\sigma_1^\prime}(\tau)\cdot W^{\sigma_2\sigma_2^\prime}(\tau)\cdot \ldots \cdot \nonumber \right. \\ && \left.  W^{\sigma_N\sigma_N^\prime}(\tau) \right)
|\sigma_1,\ldots,\sigma_N\rangle \langle\sigma_1^\prime,\ldots,\sigma_N^\prime|,
\end{eqnarray}
with the physical bonds $\sigma$ and $\sigma^\prime$. The trace is taken over virtual indices.
The size of the (virtual) dimensions of the rank-4 tensors $(W^{\sigma \sigma^\prime}_{ll^\prime})$ depends on the details of the evolution operator under consideration and will be determined for specific cases in the following.
Details on MPO representations and their practical use may be found in Ref.~\cite{SCH10}. Of course, in 1D we also
assume Hamiltonians with nearest neighbor interactions only, and the same considerations about Trotter expansions as in the 2D case apply, i.e. the time evolution will proceed in small time steps $\Delta\tau$.

Application of an MPO to an MPS will produce an MPS in terms of matrices $A^{\prime \sigma}$ with increased virtual bond dimension,
\begin{equation}
{A}^{\prime\sigma}_{(l p) (l^\prime p^\prime)}=\sum_{\sigma^\prime} W^{\sigma \sigma^\prime}_{l l^\prime} A^{\sigma^\prime}_{p p^\prime},
\end{equation}
and in order to prevent an exponential growth of the MPS size at each step of imaginary time evolution, we need to truncate the size of the MPS at
each evolution step.

\begin{figure}[h!]
    \includegraphics[width=0.35\textwidth]{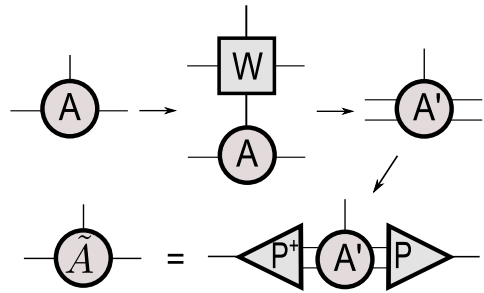}
    \vskip 3mm
    \caption{\footnotesize
    MPS renormalization }
    \label{fig:mps-update}
  \end{figure}

In order to do so one projects the MPS matrices ${A}^{\prime\sigma}$ to matrices of the same size as the original matrices $A^\sigma$.
As suggested in Ref.~\cite{PIR2010} one may use a projection operator
which is similarly constructed as in DMRG using the density  or transfer matrix,
$E=\sum_\sigma A^{\prime \sigma *} \otimes  {A}^{\prime\sigma}$. The leading eigenvector $V$ of $E$ is rewritten as
a square matrix, and from its singular value decomposition only the lowest $m$ singular values are kept. This defines
a projection operator $P$ which is used to project ${A}^{\prime\sigma}$ to a matrix $\tilde{A}^\sigma=P^\dagger {A}^{\prime\sigma} P$ of the same dimensions as the original matrix $A^\sigma$.
However, $\tilde{A}^\sigma$ corresponds to a later time of the system's evolution.
This renormalization procedure is illustrated in Fig.~\ref{fig:mps-update}.
After many imaginary-time steps and occasional reduction of the step size one reaches an approximate MPS representation of the ground state of the interacting spin system.

For some time evolution operators, MPO representations can be determined {\it exactly}.
E.g. for the interaction of a spin with an internal or external field we use the identity ($i=x,y,z$)
\begin{equation}\label{e-form1}
e^{\kappa \sigma_i}=\cosh(\kappa)\mathds{1}+\sinh(\kappa)\sigma_i.
\end{equation}
which can be proved using the properties of the Pauli matrices $\sigma_i^2=\mathds{1}$. Here $\mathds{1}$ denotes the identity matrix.

The MPO representation for the evolution operator $e^{\kappa\sum_k 1\otimes\ldots\otimes (\sigma_i)_k\otimes\ldots\otimes1}$ is
then easily obtained in terms of the $1\times 1\times 2\times 2$ tensors
\begin{equation}\label{w-matrix}
W=\Big(\cosh(\kappa)\mathds{1}+\sinh(\kappa) \sigma_i\Big),
\end{equation}
i.e. the size of virtual dimensions is 1.   We have written the $W$ tensor in terms of a variable
$\kappa=\tau g$, where $g$ is the coupling strength of the field under consideration.

For  spin-spin interactions we need the identity
\begin{equation}\label{e-form2}
e^{\kappa \sigma_i \otimes \sigma_i}= \cosh(\kappa)\mathds{1}\otimes\mathds{1}+\sinh(\kappa)(\sigma_i\otimes \sigma_i).
\end{equation}
With this relation one easily finds an MPO representation of the evolution operator $e^{\kappa\sum_k 1\otimes\ldots  (\sigma_i)_k\otimes (\sigma_i)_{k+1} \ldots\otimes 1}$ in terms of the $ 2\times 2\times 2\times 2$ $W$ tensors,
\begin{equation}
W=\left(
\begin{array}{cc}
 \cosh (\kappa) \mathds{1} & \sqrt{\sinh (\kappa) \cosh (\kappa)} {\sigma}_i \\
 \sqrt{\sinh (\kappa)\cosh (\kappa)} {\sigma}_i & \sinh (\kappa) \mathds{1} \\
\end{array}
\right),
\end{equation}
the size of the virtual dimension is 2.
The latter relation was derived using a slightly different notation in Ref.~\cite{PIR2010}.

The projection procedure for the reduction of the MPS size after each imaginary time step
is the computationally most expensive part of the calculations to be performed. Therefore, it is
desirable to streamline this step as much as possible. In fact it is desirable (from the computational viewpoint) that the $W$ tensors are real symmetric.
This then leads to a symmetric transfer operator, which can be diagonalized rather efficiently. Suitable procedures for the symmetrization of $W$ tensors are discussed in Ref.~\cite{PIR2010}.

Our present realization of the translationally invariant MPS algorithm with real symmetric
tensors $W$ is applicable only for nonnegative values of $\kappa$. In order to an have efficient algorithm for negative $\kappa$ one has to derive additional real $W$ matrices from Eq.~\eqref{e-form2} (otherwise complex numbers appear intrinsically in $W$). Instead, we use the standard TEBD algorithm~\cite{vidal:3} in our calculations for parameter dependencies which correspond to negative $\kappa$. The description of the TEBD algorithm is present in various papers (e.g.,\cite{VID2007}), thus we do not provide it in the present paper.

For the translationally invariant MPS algorithm physical quantities are calculated from the 2-spin reduced density matrix
\begin{equation}\label{redden}
\rho_{\sigma_1^\prime \sigma_2^\prime\sigma_1 \sigma_2}={\rm Tr}(\mathcal{G}\cdot
     T^{\sigma_1^\prime \sigma_1}\cdot
     T^{\sigma_2^\prime \sigma_2})
\end{equation}
in terms of the environment matrix $\mathcal{G}=(v_L\otimes v_R)^T /\lambda^2$ and the unrenormalized `impurity matrices'
$T^{\sigma^\prime \sigma}=A^{\sigma^\prime *}\otimes A^{\sigma}$.
The environment matrix is determined from the leading left $v_L$ and right $v_R$ eigenvectors of the transfer matrix $\sum_\sigma T^{\sigma \sigma}$ and its eigenvalue $\lambda$. In this way we obtain results in the thermodynamic limit (see~\cite{SCH10}). Various results calculated from this density matrix are compared with 2D results in the next section.

In the case of TEBD algorithm, a bipartition in the state representation is present, and the translationally invariant ground state is represented by two MPSs $\{A,B\}$. The 2-spin reduced density matrix is calculated as

\begin{equation}
\rho_{\sigma_1^\prime \sigma_2^\prime\sigma_1 \sigma_2}={\rm Tr}(\mathcal{G}\cdot
     T_A^{\sigma_1^\prime \sigma_1}\cdot
     T_B^{\sigma_2^\prime \sigma_2}).
\end{equation}

The impurity matrices are here:~$T_A^{\sigma^\prime \sigma}=A^{\sigma^\prime *}\otimes A^{\sigma}$,~$T_B^{\sigma^\prime \sigma}=B^{\sigma^\prime *}\otimes B^{\sigma}$. The environment matrix is~$\mathcal{G}=(v_L\otimes v_R)^T /\lambda$ with $\lambda$ the leading eigenvalue and $v_L$, $v_R$ the corresponding eigenvectors of a combined two-site transfer matrix $(\sum_{\sigma_1} T_A^{\sigma_1 \sigma_1}) \cdot (\sum_{\sigma_2} T_B^{\sigma_2 \sigma_2})$.

\section{Entanglement measures and entanglement distribution:
Numerical results and physical interpretation}\label{results}

In this section we apply the formalism presented in the previous section to 1D and 2D spin-1/2 systems: the quantum Ising model in a transverse magnetic field and the XXZ model.
We calculate various entanglement measures for these systems such as one-site entanglement entropy, one-tangle, concurrence of formation and negativity. Furthermore, we determine bounds on the localizable entanglement  in terms of the concurrence of assistance and maximal two-point correlation functions, local entanglement, and  entanglement per bond. We compare these quantities and discuss their ability to identify critical points and distinguish between different phases. Entanglement per bond is presented only for 2D models, due to the fact that our translationally invariant MPS algorithm does not provide the MPS in canonical form~\cite{vidal:3}. The mentioned entanglement measures are briefly defined in Appendix~\ref{entanglement}.

For all calculated quantities, the numerical results obtained from the TERG and CTMRG methods are nearly identical.
Differences between both methods increase slightly in the critical region, and are strongly dependent on cutting parameters used in the renormalization procedures. A more complete analysis of such issues
is under way.

An interesting characteristic we analyze using the calculated entanglement measures is the monogamy of entanglement~\cite{PhysRevA.61.052306} or -- more precisely -- the entanglement distribution between different parties. Somewhat naively, entanglement monogamy may be expressed as follows: if two parties are maximally entangled they cannot be entangled at all with a third party. Expressions for the distribution of entanglement in the form of monogamy relations for multi-qubit systems, based on the concurrence of formation $C_F$ and concurrence of assistance $C_A$ have been obtained in Refs.~\cite{PhysRevA.90.024304,COAmonogamy}. Thus, among the different entanglement measures we calculate in the present work, concurrence of formation and assistance are of primary interest.
For the models studied in the present paper, a naive entanglement monogamy analysis was done in Ref.~\cite{Syljuasen200425} using Monte-Carlo methods for the calculation of $C_F$. Here, we provide a more comprehensive analysis based on monogamy relations for $C_F$ and $C_A$.

Entanglement monogamy relations for $N$-qubit systems are obtained from the Coffman-Kundu-Wootters (CKW)~\cite{PhysRevA.61.052306} inequality,
\begin{equation}
[C_F]^2_{A|B_1B_2 \ldots B_{N-1}} \geq [C_F]^2_{AB_1} + [C_F]^2_{AB_2} + \ldots + [C_F]^2_{AB_{N-1}},
\label{CKW}
\end{equation}
where $[C_F]_{AB_i}=[C_F](\rho_{AB_i})$ is the concurrence of the reduced density matrix $\rho_{AB_i}$ and $[C_F]_{A|B_1B_2 \ldots B_{N-1}}=C(|\psi\rangle_{A|B_1B_2 \ldots B_{N-1}})$ the concurrence of the pure state $|\psi\rangle$ as defined in Ref.~\cite{PhysRevA.64.042315}. A more general form of the CKW inequality was recently suggested in Ref.\cite{PhysRevLett.113.110501}.
For N-qubit states, $C^2_{A|B_1B_2 \ldots B_{N-1}}$ can be obtained from the one-site reduced density matrix, and it is equal to the one-tangle $\tau_1$: $[C_F]^2_{A|B_1B_2 \ldots B_{N-1}}= 2 (1-\Tr\rho_A^2) = 4 \det \rho_A=\tau_1$~\cite{PhysRevA.64.042315}.

In our analysis we use two main assumptions concerning the entanglement structure of the ground states of the models we study. The first is that only the nearest neighbor concurrences give major contributions to the sum of the right hand side of \eqref{CKW}. The larger the separation between two particles the smaller is the concurrence between them. The second assumption is a consequence of the translational symmetry of the ground states and as a consequence all nearest neighbor concurrences are equal.

Taking into account these two features of the systems under consideration allows us to rewrite the inequality \eqref{CKW} for 1D and 2D models.
For 1D systems one obtains
\begin{equation}
\tau_1^{\text{1D}} \geq 2 \left[C_F^{\text{1D}}\right]^2_{nn} + \delta_F^{\text{1D}},
\end{equation}
where $\left[C_F^{\text{1D}}\right]^2_{nn}$ is the nearest neighbor concurrence of formation, and the quantity $\delta_F^{\text{1D}}$ contains all other bipartite concurrences.
Analogously in 2D one finds
\begin{equation}
\tau_1^{\text{2D}} \geq 4 \left[C_F^{\text{2D}}\right]^2_{nn} + \delta_F^{\text{2D}}.
\end{equation}
From these relations  we obtain information about the entanglement distribution in the state.
A more complete analysis of the entanglement distribution requires taking into account  next-nearest neighbor two-party and longer ranged bipartite terms in the CKW-inequality. Moreover, it is also possible to calculate three-party entanglement terms and look for their contribution to the entanglement distribution. This is numerically easily feasible for 1D systems, but is much harder for 2D models.

Dual to the CKW inequality one can derive the following relation wich involves the concurrence of assistance $C_A$ on the right hand side~\cite{PhysRevA.90.024304, COAmonogamy}
\begin{equation}
[C_F]^2_{A|B_1B_2 \ldots B_{N-1}} \leq [C_A]^2_{AB_1} + [C_A]^2_{AB_2} + \ldots + [C_A]^2_{AB_{N-1}}.
\label{coarelation}
\end{equation}
Again we introduce the quantity $\delta_A^{\text{1D}}$,
\begin{equation}
\tau_1^{\text{1D}} \leq 2 \left[C_A^{\text{1D}}\right]^2_{nn} + \delta_A^{\text{1D}},
\end{equation}
where $\left[C_A^{\text{1D}}\right]^2_{nn}$ contains the nearest neighbor terms and $\delta_A$ the longer-ranged bipartite concurrences.
In 2D we have
\begin{equation}
\tau_1^{\text{2D}} \leq 4 \left[C_A^{\text{2D}}\right]^2_{nn} + \delta_A^{\text{2D}}.
\end{equation}

\subsection{Quantum Ising model in a transverse field}

The spin-$\frac{1}{2}$ Ising model in a transverse magnetic field $h$  is given by the Hamiltonian
\begin{equation}\label{Ising Hamiltonian 1/2}
H^{\text{Ising}}=J\sum_{\langle i,j\rangle} \sigma_i^z   \otimes \sigma_{j}^z+h\sum_i \sigma_i^x,
\end{equation}
where the $\sigma_i$ $(i=1,2,3)$ are the standard Pauli spin operators. This model is $Z_2$ symmetric (spin-flip symmetric).

The sign of the coupling constant $J$ determines the type of the interaction between spins: anti-ferromagnetic for $J>0$ and ferromagnetic for $J<0$.
The calculated physical quantities are symmetric with respect to $J=0$. Quantities like the magnetization $m_x$ ($J<0$) are mapped to their staggered counterpart ($J>0$). In the present paper we choose the energy scale by setting  $J=-1$.

In 1D this model can be solved analytically using a Jordan-Wigner transformation~\cite{LIE61}.
It is well known that at the critical points $h=\pm 1$ this model shows quantum phase transitions separating a magnetically ordered
phase ($-1<h<1$) from paramagnetic phases ($h<-1$ and $h>1$). In the ordered phases the $Z_2$ symmetry is spontaneously broken.
At the critical points and in the thermodynamic limit the ground state energy per site is given by $E_0=-4/\pi$.

The 2D quantum Ising model cannot be solved analytically, and various methods are applied to solve it numerically,
notably rather resource-intensive Monte-Carlo (MC) methods.
Such calculations find a transition between a ferromagnetic and a paramagnetic phase at a critical point $h_{\rm cr}=3.044$~\cite{PhysRevE.66.066110}.
The tensor network implementation we use here produces numerical results significantly faster than MC calculations,
however, with less precision: our implementation determines a critical point at $h_{\rm cr} \approx 3.25$, which is determined from a singular point of the second derivative of the ground state energy as a function of $h$. Of course, significantly more precise results could be obtained with more elaborate tensor network implementations and larger bond sizes~\cite{orus:review}. However, it is our goal to investigate correlations and entanglement properties using small numerical cost. In practice, we study positive $h$ only and obtain results for negative $h$ by a reflection at $h=0$. In order to compare numerical results for one- and two-dimensional systems we rescale the magnetic field dependence $h/h_{\rm cr}$ such that phase transitions always occur at $h/h_{\rm cr}=1$.

For $h \ll h_c$, the system (both in 1D and 2D) is a classical Ising model with a doubly degenerate ground state (in the thermodynamic limit). In experimental situations this degeneracy is broken and this is done intrinsically in our MPS and PEPS implementations as well. For $h\gg h_{cr}$ the magnetic field dominates and the ground state corresponds to free spins oriented according to the magnetic field.

\begin{figure}[h!]
    \includegraphics[width=0.45\textwidth]{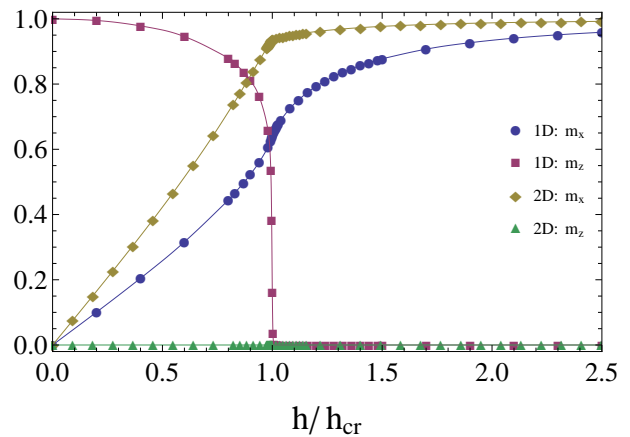}
    \caption{\footnotesize (colour online)
    Magnetizations $m_x$ and $m_z$ as a function of the magnetic field $h/h_{cr}$
    for the 1D and 2D quantum Ising models.
    Parameters for the 1D MPS calculation: $m=20$.
    Parameters for the 2D TERG calculation: $D=4$, $D_{c}=20$.
    } \label{magnet1D2D-ising}
\end{figure}

With Fig.~\ref{magnet1D2D-ising} we start the presentation of the numerical results and show the magnetizations $m_x$ and $m_z$ as a function of $h$. We see that $m_x(h/h_{\rm cr})$ in 1D increases slower from zero magnetic field towards the critical point and has lower value at the critical point then the $m_x$ in the 2D.

At this stage we do not quantitatively extract critical exponents as this would require more precise and time-consuming calculations close to the critical points. However, qualitatively the
critical properties are in agreement with expectations.

\begin{figure}[h!]
    \includegraphics[width=0.45\textwidth]{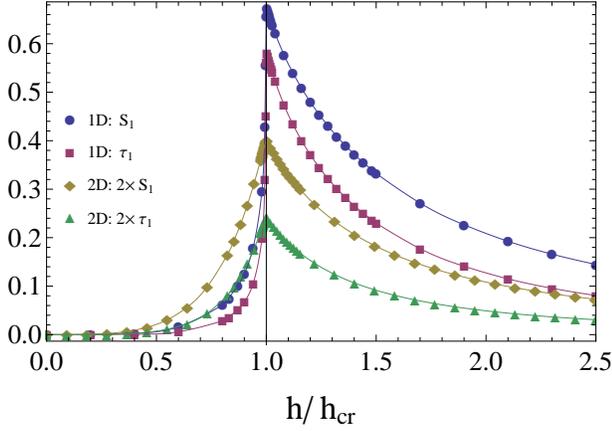}
    \caption{ \footnotesize (colour online)
    One-site entanglement entropy $S_1$ and one-tangle $\tau_1$ as a function of the magnetic field $h/h_{\rm cr}$ for the 1D and 2D quantum Ising models.
    2D results are multiplied by a factor of 2.
    Parameters for the 1D MPS calculation: $m=20$.
    Parameters for the 2D TERG calculation: $D=4$, $D_{c}=20$.
    } \label{onesite1D2D-ising}
\end{figure}

In Fig.~\ref{onesite1D2D-ising} we show the entanglement measures calculated from the single-spin density matrix:
one-site entanglement entropy $S_1$ and one-tangle $\tau_1$ for the one- and two-dimensional Ising models. It is clearly seen that all these measures nicely peak in cusps at the critical point. All 2D results are multiplied by a factor of 2 {for the convenience.}
%

\begin{figure}[h!]
    \centering
    \includegraphics[width=0.45\textwidth]{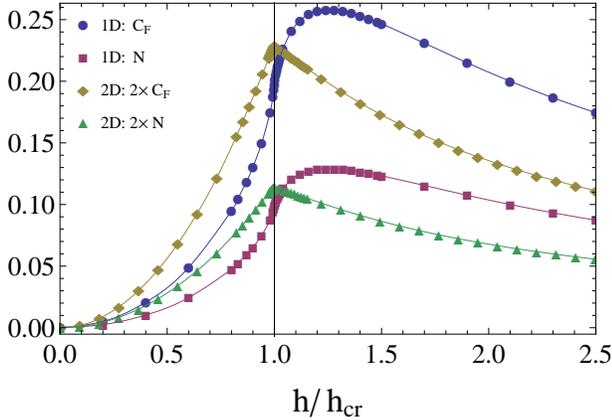}
    \caption{\footnotesize (colour online)
    Concurrence of formation $C_F$ and negativity $N$ as a function of the magnetic field $h/h_{cr}$ for 1D and 2D quantum Ising model.
    2D results are multiplied by a factor of 2.
    Parameters for the 1D MPS calculation: $m=20$.
    Parameters for the 2D TERG calculation: $D=4$, $D_c=20$.
    } \label{concFandNeg1D2D-ising}
\end{figure}

\begin{figure}[h!]
    \includegraphics[width=0.45\textwidth]{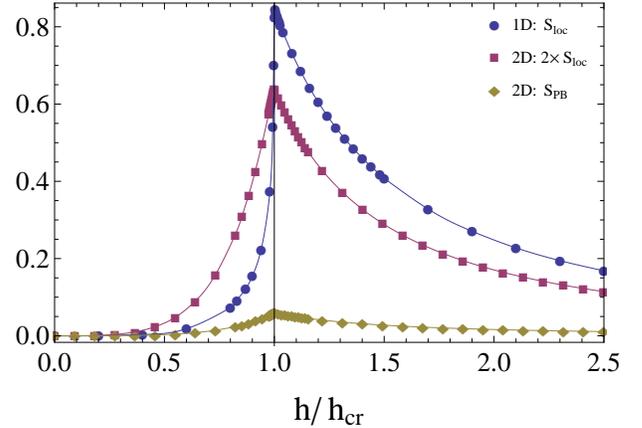}
    \caption{\footnotesize (colour online)
    Comparison of local entanglement $S_{\rm loc}$ dependence on magnetic field $h$ for the 1D and 2D quantum Ising model. Entanglement per bond $S_{PB}$ dependence for the 2D Ising model.
    Results are renormalized to the $h/h_{cr}$ dependence.
    Results for the 2D model are multiplied by a factor of 2.
    Parameters for the 1D MPS calculation: $m=20$.
    Parameters for the 2D TERG calculation: $D=4$, $D_c=20$.
    }\label{sLocandSPB1D2D-ising}
\end{figure}

In Figs.~\ref{concFandNeg1D2D-ising} and~\ref{sLocandSPB1D2D-ising} we show entanglement measures calculated from the two-spin density matrix as a function of the magnetic field: the concurrence of formation $C_F$ and negativity $N$.
Our TN results are in a very good agreement with the Monte Carlo results by Syljuasen~\cite{Syljuasen200425}.
Of course, calculations close to the critical point {in 2D} are difficult both for TN and MC methods, but clearly the cusp at the critical point can be better resolved with the TN method used here. Close to the critical point the MC results of Ref.~\cite{Syljuasen200425} are very noisy.
The concurrence of formation for 1D does not peak  at the critical point, but shows an inflection. This is in agreement with the analytical results presented in Ref.~\cite{Osterloh2002}.

The negativity shows similar characteristics as the concurrence of formation both in 1D and 2D. Concurrence of formation and negativity reach their maximum at the same value for the magnetic field.
Negativities for 1D and 2D geometries both satisfy the concurrence bounds $\sqrt{(1-C_F)^2+C_F^2}-(1-C_F)\leq N \leq C_F$~\cite{0305-4470-34-47-329}.

The local entanglement $S_{\rm loc}$  shown in Fig.~\ref{sLocandSPB1D2D-ising} is the simplest form of a block-block entanglement, the entanglement between two neighboring spins and their environment. We see that both in 1D and 2D this measure has a peak {with a cusp} at the critical point. Not surprisingly, {$S_{\rm loc}$'s absolute value at the critical point is the largest among other entanglement measures we calculate from the two-site reduced density matrix. This is due to the fact that $S_{\rm loc}$ correspond to entanglement between two neighbor spins as one party with all other spins as another party in contrast to the entanglement between just two neighbor spins in the case of $C_F$, $N$, $C_A$.}
Similar to the one-site entanglement entropy, $S_{\rm loc}$ is small in the paramagnetic phase ($h<h_{cr}$), increases sharply close to the critical point and then decreases slowly in the ferromagnetic phase ($h>h_{cr}$).

Fig.~\ref{sLocandSPB1D2D-ising} also demonstrates that the bipartite entanglement per bond in 2D identifies the critical point {having a peak with a cusp there}. This measure exemplifies one useful advantage of the translationally invariant TN methods: the possibility to extract information about the state right from the TN representation, i.e.  one does not need to calculate expectation values at potentially high numerical cost.

\begin{figure}[h!]
    \includegraphics[width=0.45\textwidth]{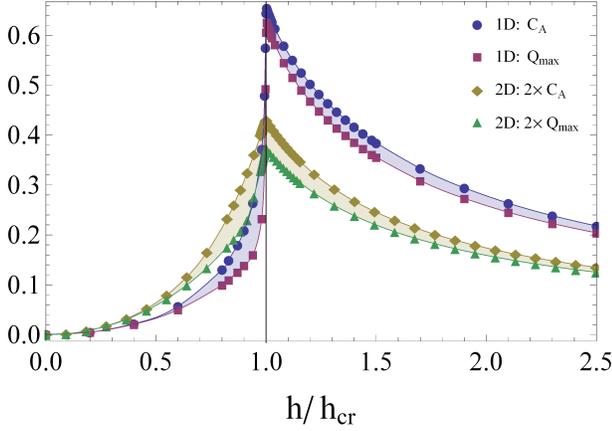}
    \caption{\footnotesize (colour online)
    Upper and lower bounds of the localizable entanglement
    as a function of magnetic field $h/h_{cr}$ for the 1D and 2D quantum Ising models;
    2D results are multiplied by a factor of 2.
    The shaded areas between $C_A$ and $Q_{\text{max}}$ for 1D and 2D results correspond to possible values of the localizable entanglement.
    Parameters for the 1D MPS calculation: $m=20$.
    Parameters for the 2D TERG calculation: $D=4$, $D_{c}=20$.
    }\label{sLocaliz1D2D-ising}
\end{figure}

In Fig.~\ref{sLocaliz1D2D-ising} we compare the upper bound {(concurrence of assistance $C_A$)} and lower bound {(maximal two-site correlation function $Q_{\text{max}}$)} of the localizable entanglement as a function of the magnetic field. Our results show cusps at the critical point and are in a good agreement with those obtained using other methods~\cite{PhysRevLett.92.027901, Syljuasen200425}.
Note, that our MPS and PEPS implementation intrinsically break the $Z_2$ symmetry,
thus leading to product states for small and large magnetic fields.


All entanglement measures discussed above are able to identify the critical point of the system both in 1D and 2D. 
The fastest and easiest way to identify the critical point is obtained from the entanglement per bond.
This measure explicitly requires a tensor network representation and cannot be obtained using other methods.
As expected, all entanglement measures approach zero for small and large transverse magnetic fields, which indicates product states for these limits.

\begin{figure}[h!]
    \centering
    \includegraphics[width=0.45\textwidth]{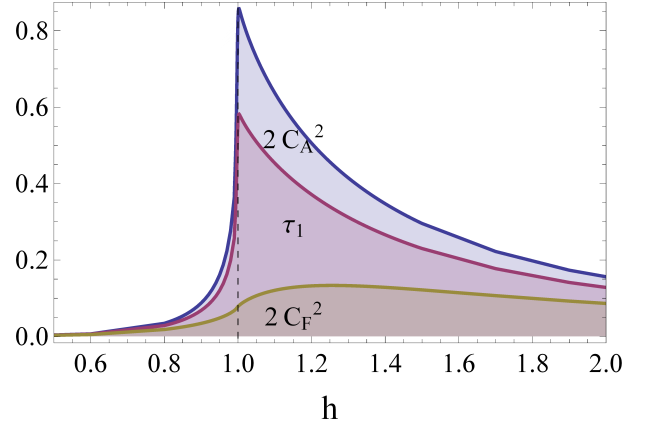}
    \caption{\footnotesize (colour online)
    Entanglement monogamy analysis for the 1D quantum Ising model:
    Comparison of the concurrence of formation $C_F$, the concurrence of
    assistance $C_A$ and the 1-tangle $\tau_1$.
    For details see the discussion in the main text.
    Parameters for the 1D MPS calculation: $m=20$.
    }
    \label{monogamy1-ising}
\end{figure}

In Fig.~\ref{monogamy1-ising} we show the concurrence of assistance $2\left[C_A^{\text{1D}}\right]^2_{nn}$, the one-tangle $\tau_1^{\text{1D}}$, and the concurrence of  formation $2\left[C_F^{\text{1D}}\right]^2_{nn}$. Comparing $\tau_1^{\text{1D}}$ and $2\left[C_F^{\text{1D}}\right]^2_{nn}$ we see that the CKW inequality~(\ref{CKW}) is fulfilled and that the nearest-neighbor two-particle entanglement corresponds to only about $25\%$ of the bipartite entanglement in the critical region. At the same time, outside of the critical region $2\left[C_F^{\text{1D}}\right]^2_{nn}$ nearly exhausts the CKW inequality. This behaviour quantitatively confirms that the phase transition is characterized by the presence of long-range entanglement.
Comparing $\tau_1^{\text{1D}}$ and $(2\left[C_A^{\text{1D}}\right]^2_{nn})$ we conclude that already the nearest neighbor entanglement contributions are larger than the lower bound $\tau_1^{\text{1D}}$ of how much entanglement can be created by assistance.

\begin{figure}[h!]
    \centering
    \includegraphics[width=0.45\textwidth]{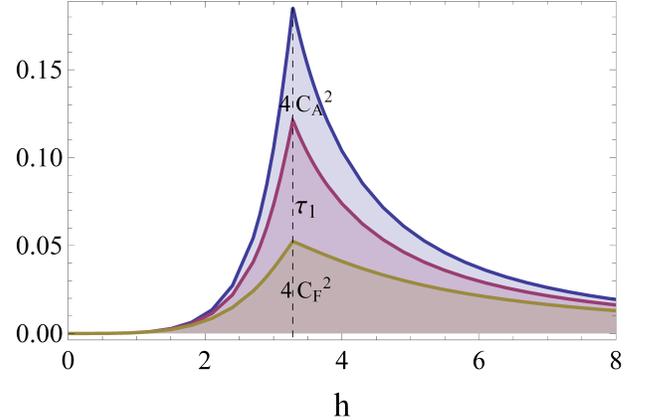}
    \caption{\footnotesize (colour online)
    Entanglement monogamy analysis for the 2D quantum Ising model:
    Comparison of the concurrence of formation $C_F$, the concurrence of
    assistance $C_A$ and the 1-tangle $\tau_1$.
    For details see the discussion in the main text.
    Parameters for the 2D TERG calculation: $D=4$, $D_{c}=20$. Critical value of the magnetic field is $h_{\rm cr}\approx 3.28$.
    }
    \label{monogamy2-ising}
\end{figure}

Fig.~\ref{monogamy2-ising} displays the entanglement monogamy analysis for the 2D quantum Ising model. Here, we compare $4\left[C_A^{\text{2D}}\right]^2_{ij}$, $\tau_1^{\text{2D}}$ and $4\left[C_F^{\text{2D}}\right]^2_{ij}$. The CKW inequality is fulfilled, and the nearest-neighbor  entanglement in the critical region corresponds  to about $50\%$ total bipartite entanglement. In comparison to the 1D result, we observe that the 2D
nearest-neighbor entanglement contains more of the total bipartite  entanglement, which can be explained by the presence of a larger number of nearest neighbors of each site. Again, similar to the 1D case, $4\left[C_F^{\text{1D}}\right]^2_{nn})$ nearly exhausts the CKW inequality outside of the critical region.
Again, comparing $\tau_1^{\text{2D}}$ and $4\left[C_A^{\text{1D}}\right]^2_{ij}$ we see that also in 2D the nearest-neighbor entanglement terms in general are already larger than the lower bound $\tau_1^{\text{2D}}$ on how much entanglement can be created by assistance.

\subsection{XXZ model}
Next we study the spin-$\frac{1}{2}$ XXZ (anisotropic Heisenberg) model,
\begin{equation}\label{Ising Hamiltonian}
H^{\text{XXZ}}=\sum_{\langle i,j \rangle} \left\{ \sigma_i^x \otimes\sigma_j^x + \sigma_i^y \otimes\sigma_j^y + \Delta \, \sigma_i^z \otimes\sigma_j^z \right\}
\end{equation}
as a function of the anisotropy parameter $\Delta$.
The Hamiltonian of this model is ${\rm U}(1)$ symmetric (corresponing to an invariance under a ${\rm U}(1)$ rotation about the spin $z$ axis) as well as $Z_2$ symmetric (corresponding to an invariance under a  $\pi$ rotation about the spin $x$ or $y$ axis). It is ${\rm SU(2)}$ symmetric at the Heisenberg point $\Delta=1$.
The ground state of the XXZ model in different phases  preserves these symmetries or not depending of the space dimension.~\cite{PhysRevA.68.060301}.
The $Z_2$ symmetry implies that $\langle \sigma^z_i \rangle = 0$ and $\langle\sigma^x_i\sigma^z_j\rangle=\langle\sigma^y_i\sigma^z_j\rangle=0$. The ${\rm U}(1)$ symmetry implies that
$\langle \sigma^x_i \rangle = \langle \sigma^y_i \rangle= 0$, $\langle\sigma^x_i\sigma^x_j\rangle = \langle\sigma^y_i\sigma^y_j\rangle$, $\langle\sigma^x_i\sigma^y_j\rangle =0$.

The XXZ model has a richer phase structure than the Ising model:
The 1D XXZ model shows three phases~\cite{Mikeska01,PhysRevA.85.052128}.
For $\Delta>1$ the system is in a gapped antiferromagnetic phase (in particular, it corresponds to a classical Ising anti-ferromagnet
for large positive $\Delta$). At $\Delta = 1$ there is a critical point, where an infinite-order Kosterlitz-Thouless quantum phase transition occurs
from the anti-ferromagnetic phase to the XY phase. The system is equivalent here to the spin-$\frac{1}{2}$ Heisenberg anti-ferromagnet with a gapless ground state.
In the XY phase ($|\Delta|<1$) the system is gapless and the correlation functions decay polynomially. At $\Delta=-1$ the system undergoes a
first-order quantum phase transition to a ferromagnetic gapped phase for $\Delta \leq -1$. For large negative $\Delta$ the system resembles an Ising ferromagnet.

In the thermodynamic limit spontaneous $Z_2$ symmetry breaking occurs in the ferromagnetic ($\Delta<-1$) and antiferromagnetic phases ($\Delta>1$), but $Z_2$ symmetry is preserved in the XY phase. The continuous ${\rm U}(1)$ symmetry remains unbroken in all three phases of the 1D XXZ model.

The two-dimensional XXZ model shows three different phases, as well~\cite{PhysRevB.49.9702,PhysRevB.65.092402,PhysRevB.64.214411}: an antiferromagnetic phase for $\Delta>1$, an XY phase for $|\Delta|<1$ and a ferromagnetic phase for $\Delta<-1$. It undergoes a second-order phase transition at $\Delta=1$~\cite{PhysRevB.84.174426} and a first-order phase transition at $\Delta=-1$~\cite{0953-8984}. Just as in 1D, the $Z_2$ symmetry is spontaneously broken in the ferromagnetic ($\Delta<-1$) and antiferromagnetic phases ($\Delta>1$), and remains unbroken in the XY phase. However, unlike in the 1D case, the continuous ${\rm U}(1)$ symmetry is not preserved in the XY phase of the 2D XXZ model~\cite{PhysRevA.68.060301}.

The one-dimensional XXZ model has been studied extensively using the Bethe Ansatz~\cite{PhysRev.150.321,PhysRev.150.327,PhysRev.151.258,mattis:book}. At the Heisenberg point ($\Delta=1$) the ground state energy is found to be $E_0= -4\log 2+1$.
In order to demonstrate the dependence of our numerical results on the MPS virtual dimension, in Table~\ref{XXZ1D_tab} we compare the ground state energy calculated for different $m$ with the analytical value.
In the following we will present  results calculated with $m=20$ which provides accurate results at moderate numerical costs. For the $\Delta<0$ region we will use the TEBD imaginary time evolution method instead of the translationally invariant MPS method, because our MPS implementation is not optimal for calculations in this region by construction. Tests with both algorithms in the region $\Delta>0$ show that the  results agree to a high precision.

\begin{table}
\centering
\begin{tabular}{l l l l l l r}
 m            & $E$       & $\Delta E $     \\
 \hline
 10           & -1.77202 &  1.0 $10^{-3}$  \\
 15           & -1.77237 &  3.8 $10^{-4}$  \\
 20           & -1.77247 &  2.1 $10^{-4}$  \\
 25           & -1.77253 &  1.1 $10^{-4}$  \\
 30           & -1.77254 &  8.5 $10^{-5}$  \\
\hline
 BA           & -1.77259 &                &         &         &            &         \\
\end{tabular}
  \caption{\footnotesize Ground state energy $E_0$ of the 1D XXX model compared to the Bethe Ansatz (BA) result as a function of MPS virtual dimension $m$.
  The relative difference is $\Delta E=(E_0-E_{\rm BA})/E_{\rm BA}$.
  \label{XXZ1D_tab}}
\end{table}

The Heisenberg point ($\Delta=1$) for the 2D XXZ model was intensively studied in Ref.~\cite{gu:1,PhysRevB.56.11678}. The best quantum Monte Carlo result for the ground state energy is $E^{QMC}=-1.340$~\cite{PhysRevB.56.11678}. In Table~\ref{XXZ2D_tab} we compare our numerical results for this energy for different $D$ and $D_c$.
In the following we will present results calculated using $D=4$ and $D_{c}=20$. With this choice we obtain good results at reasonable numerical cost.
\begin{table}
\centering
\begin{tabular}{l l l l l l r}
 $D$ & $D_{c}$      & $E$       & $\Delta E $     \\
 \hline
 2    & 20           & -1.318 &  2.2 $10^{-2}$  \\
 3    & 20           & -1.327 &  1.3 $10^{-2}$  \\
 4    & 32           & -1.333 &  7.0 $10^{-3}$  \\
 5    & 64           & -1.338 &  2.0 $10^{-3}$  \\
\hline
 QMC  &              & -1.340 &                  \\
\end{tabular}
  \caption{\footnotesize Ground state energy $E_0$ of the 2D XXX model compared to QMC results as a function of PEPS virtual dimension $D$ and TERG cutting parameter $D_c$. The relative difference is $\Delta E=(E_0-E_{\rm QMC})/E_{\rm QMC}$.
  \label{XXZ2D_tab}}
\end{table}

\begin{figure}[h!]
    \centering
    \includegraphics[width=0.45\textwidth]{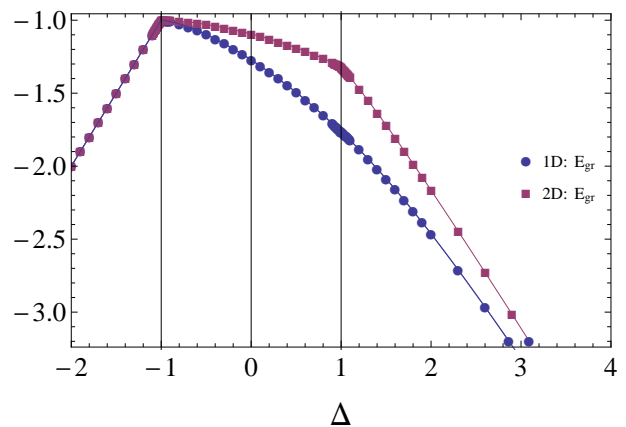}
    \caption{\footnotesize (color online)
    Comparison of ground state energy as a function of $\Delta$ for the 1D and 2D XXZ models.
    Parameters for the 1D MPS calculation: $m=20$.
    Parameters for the 2D TERG calculation: $D=4$, $D_{c}=20$.
    }
    \label{energy1D2D-xxz}
\end{figure}

In  Fig.~\ref{energy1D2D-xxz} we show the ground state energy per site as a function of the asymmetry parameter $\Delta$ for 1D and 2D.
From the figure we see that in the parameter region $\Delta < -1$ the ground state energy is linearly dependent on $\Delta$: $E_0(\Delta)=\Delta$ for both the 1D and 2D XXZ models.
At $\Delta=1$ the ground energy shows a kink for 2D, but for 1D is continuous and infinite-order differentiable as is known from analytical analysis~\cite{PhysRev.150.321,PhysRev.150.327,PhysRev.151.258}.

\begin{figure}[h!]
    \centering
    \includegraphics[width=0.45\textwidth]{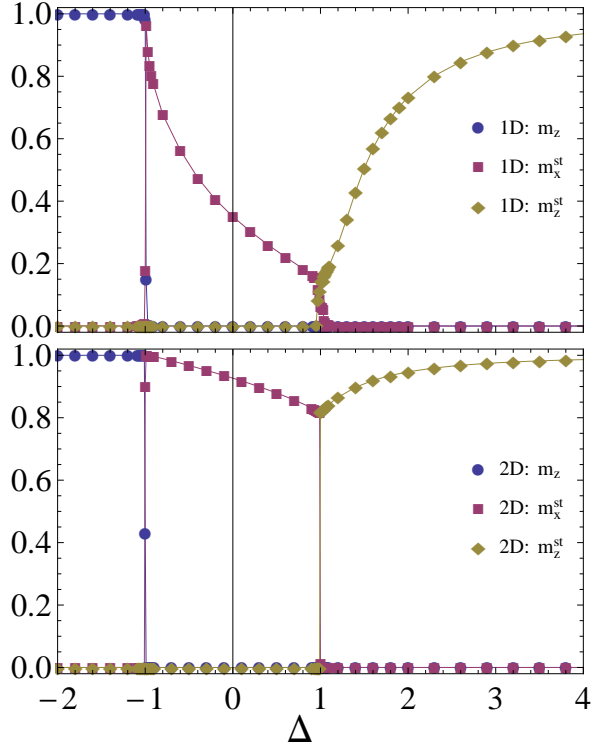}
    \caption{ \footnotesize (colour online)
    Comparison of magnetization $m_z$ and its staggered counterpart $m_z^{\text{st}}$ as a function of $\Delta$ for the 1D and 2D XXZ models.
    ${\rm U}(1)$ symmetry breaking for $|\Delta|<1$ also gives nonzero staggered  magnetization $m_x^{\text{st}}$.
    Parameters for the 1D MPS calculation: $m=20$.
    Parameters for the 2D TERG calculation: $D=4$, $D_{c}=20$.
    }
    \label{magnet1D2D-xxz}
\end{figure}

Fig.~\ref{magnet1D2D-xxz} shows various magnetizations as a function of the asymmetry parameter $\Delta$.
Non-zero $m_z$ magnetization in the ferromagnetic phase ($\Delta<-1$) and non-zero staggered magnetization $m_z^{\text{st}}$ in the
anti-ferromagnetic phase ($\Delta>1$) both in 1D and 2D models confirm the $Z_2$ symmetry breaking in these phases. In the XY phase, Fig.~\ref{magnet1D2D-xxz} shows a ${\rm U}(1)$ symmetry breaking not only for the 2D model,
as expected, but also for the 1D model. This clearly shows a deficiency of the numerical method used. This ${\rm U}(1)$ symmetry breaking is strongly dependent on the chosen $m$,
and gets smaller with increasing $m$, however, it appears that one needs to use a code which implements the U(1) symmetry of the states from the outset in order to get more precise results. We will do this in a future paper.
This unphysical breaking of the U(1) symmetry will be seen as well later in various calculated entanglement measures. Note that the translationally invariant MPS algorithm nicely obtains the anti-ferromagnetic phase, despite the fact that it uses equal tensors at each site.
The magnetizations in one dimension are weaker than their two-dimensional counterparts.

\begin{figure}[h!]
    \centering
    \includegraphics[width=0.45\textwidth]{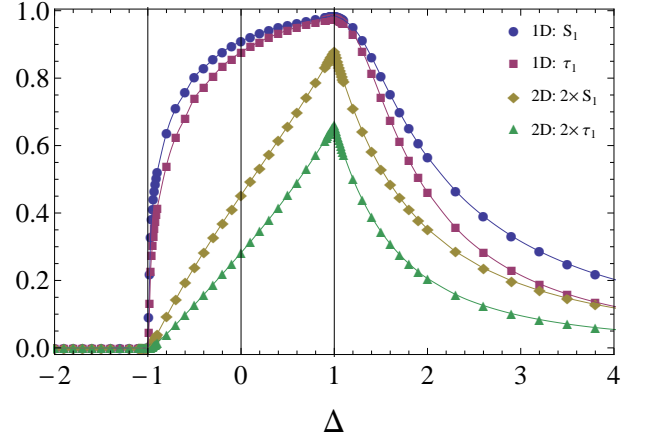}
    \caption{\footnotesize (colour online)
    Comparison of one-site entanglement entropy $S_1$ and one-tangle $\tau_1$ dependence on $\Delta$ for the 1D and 2D XXZ models.
    Results for 2D model are  multiplied by a factor of 2.
    Parameters for the 1D MPS calculation: $m=20$.
    Parameters for the 2D TERG calculation: $D=4$, $D_{c}=20$.
    }
    \label{onesite1D2D-xxz}
\end{figure}

In Fig.~\ref{onesite1D2D-xxz} we show the one-site entanglement measures: one-site entanglement entropy $S_1$ and one-tangle $\tau_1$ for the one- and two-dimensional XXZ models.
All these measures  peak in  cusps at the critical point $\Delta=1$ and are zero for the $\Delta<-1$.
At the Heisenberg point in the 1D model the ground state is SU(2) symmetric and the one-site measures $S_1$ and $\tau_1$ approach their maximal possible values.
Theoretically it is expected that these quantities equal to 1 throughout the XY phase in 1D, but due to the ${\rm U}(1)$ symmetry breaking introduced
in the algorithms as discussed above these quantities decrease while approaching the $\Delta=-1$ critical point. Again in 1D we would obtain better results for larger $m$ or by using a code which respects the U(1) symmetry from the outset.

\begin{figure}[h!]
    \centering
    \includegraphics[width=0.45\textwidth]{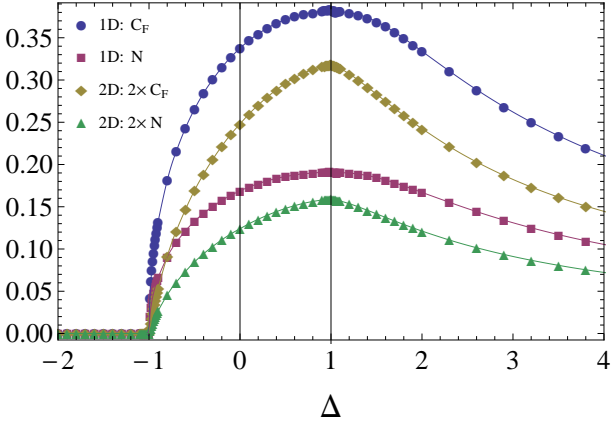}
    \caption{\footnotesize (colour online)
    Concurrence of formation $C_F$ and negativity $N$ as a function of $\Delta$ for the 1D and 2D XXZ models.
    Results for the 2D model are multiplied by a factor of 2.
    Parameters for the 1D MPS calculation: $m=20$.
    Parameters for the 2D TERG calculation: $D=4$, $D_{c}=20$.
    }
    \label{concFandNeg1D2D-xxz}
\end{figure}

In Fig.~\ref{concFandNeg1D2D-xxz} we show the two-site entanglement measures: concurrence of formation $C_F$ and negativity $N$ for the one- and two-dimensional XXZ models.
The concurrence of formation for the 1D and 2D XXZ models was studied in Refs.~\cite{PhysRevA.68.060301,Syljuasen200425,PhysRevA.71.052322}, and our results are in a very good agreement. The figure nicely shows that $C_F$ and $N$ in one and two dimensions have maxima exactly at the critical point $\Delta=1$.
It is known that $C_F$ is related to the ground state energy~\cite{PhysRevLett.93.250404}.
We see that similarly to the ground state energy in Fig.~\ref{energy1D2D-xxz}, $C_F$ and $N$ show a maximum at the critical point.
Our results correspond to the fact discussed in Ref.~\cite{PhysRevB.64.214411} that the ground state energy of the XXZ model in two and three dimensions shows a cusp at the transition point, {thus leading to a cusp in concurrence of formation. The 1D  $C_F$ and $N$ just have maxima at the  critical point $\Delta=1$ without cusps.}

Negativity for the XXZ model was previously studied for a two-qubit chain~\cite{Meng01} and for infinite tree tensor network states~\cite{refId0}. Our results extend such studies to infinite chains and infinite square-lattice systems. Again, negativities for 1D and 2D geometries both satisfy the concurrence bounds $\sqrt{(1-C_F)^2+C_F^2}-(1-C_F)\leq N \leq C_F$~\cite{0305-4470-34-47-329}.
Similarly to the quantum Ising model, we find that negativities for the 1D and 2D XXZ models have a similar behavior as the concurrence of formation in 1D and 2D.
{The 2D negativity peaks in a cusp and the 1D negativity just shows maximum at the critical point $\Delta=1$.}


\begin{figure}[h!]
    \centering
    \includegraphics[width=0.45\textwidth]{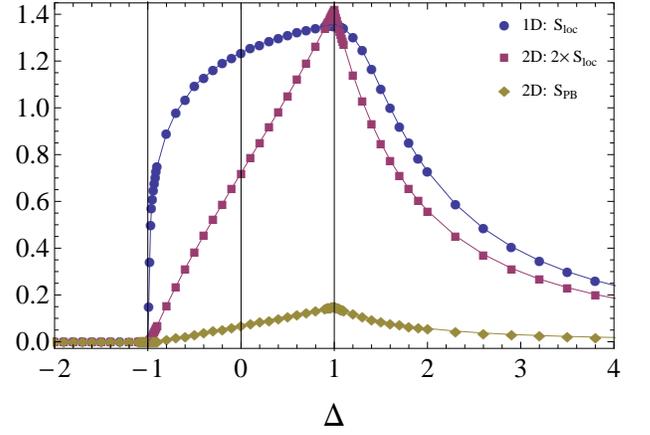}
    \caption{\footnotesize  (colour online)
    Local entanglement $S_{\rm loc}$ and entanglement per bond $S_{PB}$ as a function of $\Delta$ for 1D and 2D XXZ model.
    Results for 2D model are  multiplied by a factor of 2.
    Parameters for MPS calculation: $m=20$.
    Parameters for TERG calculation: $D=4$, $D_{c}=20$.
   }
    \label{sLocandSPB1D2D-xxz}
\end{figure}

In Fig.~\ref{sLocandSPB1D2D-xxz} we present the  local entanglement $S_{\rm loc}$ for the one- and two-dimensional XXZ models. The entanglement per bond $S_{PB}$ for the 2D XXZ model is also shown. Local entanglement for the 2D XXZ model was studied in~\cite{1367-2630-8-4-061}. However, $S_{\rm loc}$ requires comment: for $\Delta \gg 1$ we observe that the local entanglement reported here approaches zero while in Ref.~\cite{GuTianLin2006} it approaches 1.
The reason for this difference is the fact that the ground states we consider here has broken $Z_2$ symmetry, while the authors of Ref.~\cite{GuTianLin2006}  assume that the ground state is $Z_2$ symmetric.
{We see that local entanglement for 1D and 2D has similar behaviour as the one-site entanglement entropy $S_1$.}
{In both 1D and 2D $S_{\rm loc}$ vanishes at $\Delta=-1$ and peaks in a cusp at $\Delta=1$.}

Entanglement per bond for the 2D XXZ model was analyzed in~\cite{PhysRevA.81.032304}, but the authors discuss the $S_{PB}$ dependence on an external magnetic field with some fixed $\Delta$. In our studies we have no external magnetic field and vary the anisotropy parameter $\Delta$. Similarly to the quantum Ising model, $S_{PB}$ shows its ability to determine critical points {by vanishing at $\Delta=-1$ and having a peak with a cusp at $\Delta=1$.}

\begin{figure}[h!]
    \centering
    \includegraphics[width=0.45\textwidth]{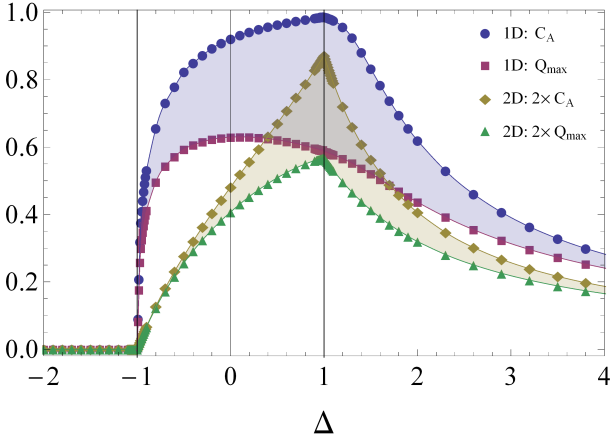}
    \caption{\footnotesize (colour online)
    Bounds on the localizable entanglement as a function of $\Delta$ for the 1D and 2D XXZ models.
    Results for the 2D model are  multiplied by a factor of 2.
    The shaded areas between $C_A$ and $Q_{\text{max}}$ for 1D and 2D results correspond to possible values of the localizable entanglement.
    Parameters for the 1D MPS calculation: $m=20$.
    Parameters for the 2D TERG calculation: $D=4$, $D_{c}=20$.
    }
    \label{sLocaliz1D2D-xxz}
\end{figure}

In Fig.~\ref{sLocaliz1D2D-xxz} we show the upper bound {(concurrence of assistance $C_A$)} and lower lower {(maximal two-site correlation function $Q_{\text{max}}$)}  on the localizable entanglement for the one- and two-dimensional XXZ models. These bounds in the two-dimensional case were also studied in Ref.~\cite{Syljuasen200425}.

We observe that for $|\Delta|<1$  the concurrence of assistance $C_A$ and two-point correlation function $Q_{\text{max}}$ decrease for smaller $\Delta$ while in Ref.~\cite{Syljuasen200425} $C_A=1$ throughout the XY phase and $Q_{\text{max}}$ does not drop to zero at one of the critical points. This difference in our results and results from~\cite{Syljuasen200425} can be explained as following.
It was shown in Ref.~\cite{PhysRevA.68.060301} that concurrence of formation $C_F$ is unaffected by spontaneous symmetry breaking (namely, ${\rm U}(1)$ symmetry breaking) for the zero-field XXZ-model.
Let us consider also the concurrence of assistance $C_A$. The formula for $C_A$ for maintained ${\rm U}(1)$ symmetry and broken $Z_2$ symmetry was introduced in~\cite{Syljuasen200425}:
\begin{equation}
\begin{split}
C_A& =\frac{1}{2}\sqrt{(1+\langle \sigma_i^z\sigma_j^z\rangle )^2 - \langle \sigma_i^x+\sigma_j^x\rangle^2} \\
& +\frac{1}{2}\sqrt{(1-\langle \sigma_i^z\sigma_j^z\rangle )^2 - \langle \sigma_i^x-\sigma_j^x\rangle^2}.
\end{split}
\label{coaxy}
\end{equation}
Following the ideas from Ref.~\cite{PhysRevA.68.060301} for deriving the expression for $C_F$ for broken ${\rm U}(1)$ symmetry and maintained $Z_2$ symmetry, we find that $C_A$ is in this case
\begin{equation}
C_A=\frac{1}{2}\left( \sqrt{(1+\langle \sigma_i^x\sigma_j^x\rangle)^2 -4 \langle \sigma_i^x\rangle^2} +1- \langle \sigma_i^x\sigma_j^x\rangle \right).
\label{coaantiferro}
\end{equation}

Obviously, $C_A$ (unlike $C_F$) is affected by U(1) symmetry breaking.

When ${\rm U}(1)$ and $Z_2$ symmetries are obeyed, $C_A=1$. This is theoretically predicted, e.g., for the Heisenberg point $\Delta=1$. We see from our 1D results that indeed $C_A(\Delta=1)\approx 1$. {At the same time our 2D results for $C_A$ for $\Delta=1$ do not reach the value $C_A=1$. This can be explained by the fact that it is numerically hard to converge to the point where both $\langle \sigma_i^x\rangle$ and $\langle \sigma_i^z\rangle$ are zero, thus giving $C_A=1$ from both equations} \eqref{coaxy} and \eqref{coaantiferro}.

The discrepancy of our result for $Q_{\text{max}}$ and the corresponding result from Ref.~\cite{Syljuasen200425} can be explained by  ${\rm U}(1)$ symmetry breaking, resulting in a nonzero $\langle \sigma^x_i \rangle$. While $\langle \sigma^x_i \rangle$ increases,
the function $Q^{xx}=\langle \sigma_i^x\sigma_j^x\rangle-\langle \sigma^x_i \rangle\langle \sigma^x_j \rangle$ (which is larger than $Q^{yy}$ and $Q^{zz}$ in the XY phase) decreases to zero.

Thus, we see that all entanglement measures discussed above are zero for $\Delta<-1$ and also approach zero for large positive $\Delta$, indicating a product state in this limit.

\begin{figure}[h!]
    \centering
    \includegraphics[width=0.45\textwidth]{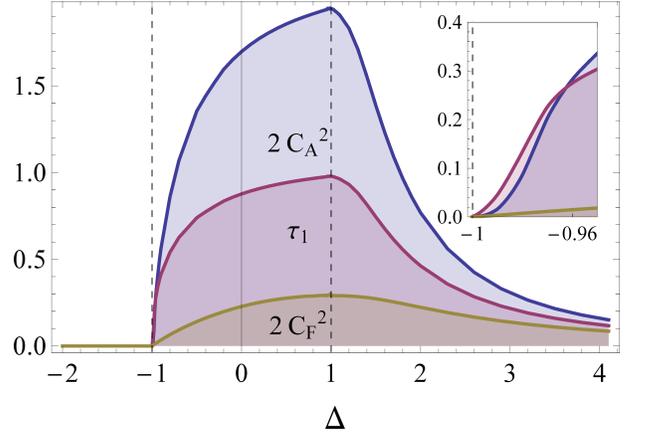}
    \caption{\footnotesize (colour online)
    Entanglement monogamy analysis for the 1D XXZ model:
    Comparison of the concurrence of formation $C_F$, the concurrence of
    assistance $C_A$ and the 1-tangle $\tau_1$.
    For details see the discussion in the main text.
    Parameters for the 1D MPS calculation: $m=20$.
    }
    \label{monogamy1-xxz}
\end{figure}

For the monogamy analysis in the Fig.~\ref{monogamy1-xxz} we represent nearest neighbor entanglement, given by concurrence of assistance  $(2\left[C_A^{\text{1D}}\right]^2_{nn})$, one-tangle $\tau_1^{\text{1D}}$ and nearest neighbor entanglement, given by concurrence of formation $(2\left[C_F^{\text{1D}}\right]^2_{nn})$. By comparing $\tau_1^{\text{1D}}$ and $(2\left[C_F^{\text{1D}}\right]^2_{nn})$ we see that CKW inequality is fulfilled and the nearest neighbor two-particle entanglement corresponds to about $1/3$ fraction of the entanglement in the critical region around $\Delta=1$ critical point. For large $\Delta\gg 1$ the nearest neighbor two-particle entanglement approaches $\tau_1^{\text{1D}}$.

By comparing $\tau_1^{\text{1D}}$ and $2\left[C_A^{\text{1D}}\right]^2_{nn}$ we see that nearest neighbor entanglement in general is larger than the lower bound $\tau_1^{\text{1D}}$ on how much entanglement can be created by assistance. Only in the XY phase in the region close to $\Delta=-1$ the nearest neighbor entanglement does not exceed the $\tau_1^{\text{1D}}$. This feature is shown in the inset  in the Fig.~\ref{monogamy1-xxz}.

\begin{figure}[h!]
    \centering
    \includegraphics[width=0.45\textwidth]{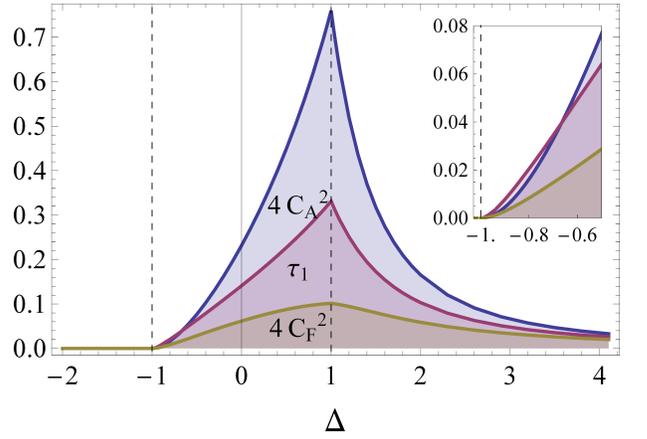}
    \caption{\footnotesize (colour online)
    Entanglement monogamy analysis for the 2D XXZ model:
    Comparison of the concurrence of formation $C_F$, the concurrence of
    assistance $C_A$ and the 1-tangle $\tau_1$.
    For details see the discussion in the main text.
    Parameters for the 2D TERG calculation: $D=4$, $D_{c}=20$.
    }
    \label{monogamy2-xxz}
\end{figure}

Fig.~\ref{monogamy2-xxz} shows the entanglement monogamy analysis for the 2D XXZ model. In this case we compare $(4\left[C_A^{\text{2D}}\right]^2_{ij})$, $\tau_1^{\text{2D}}$ and $(4\left[C_F^{\text{2D}}\right]^2_{ij})$. The CKW inequality is fulfilled. Nearest neighbor two-particle entanglement is less then $1/3$ fraction of the entanglement in the entanglement distribution in the critical region around $\Delta=1$ critical point. Similar to 1D case, for high $\Delta\gg 1$ nearest neighbor two-particle entanglement approaches $\tau_1^{\text{2D}}$.

By comparing $\tau_1^{\text{2D}}$ and $4\left[C_A^{\text{1D}}\right]^2_{nn}$ we see that nearest neighbor entanglement in general is larger than the lower bound $\tau_1^{\text{2D}}$ on how much entanglement can be created by assistance. And again, only in the XY phase in the region close to $\Delta=-1$ the nearest neighbor entanglement does not exceed $\tau_1^{\text{1D}}$,  which is shown in the inset in the Fig.~\ref{monogamy2-xxz}.

\section{Conclusions}\label{conclu}
We have investigated entanglement properties of infinite 1D  and 2D spin-1/2 systems using tensor network methods: the Ising model in transverse field and the XXZ model.
Specifically we used a translationally invariant MPS method in 1D and TERG and CTMRG in 2D in order to calculate the ground state of those models.
Different entanglement measures, such as one-site entanglement entropy and one-tangle, concurrence of formation and negativity, bounds on localizable entanglement (concurrence of assistance and two-point correlation function),
local entanglement and entanglement per bond were calculated.

Many of our results are in good agreement with those obtained using other numerical methods. This agreement underlines that such  tensor network methods are powerful tools for the investigation of quantum models.
The translationally invariant MPS algorithm and the TEBD algorithm lead to an ${\rm U}(1)$ symmetry breaking in the XY phase for the 1D XXZ model and, therefore, our results are at variance with those assuming U(1) symmetry~\cite{Mikeska01,PhysRevA.85.052128}.

Our results confirm the observation~\cite{PhysRevA.81.032304} that the bipartite entanglement per bond can successfully determine critical points. This measure is unique to tensor network methods.

We made an entanglement monogamy analysis: The Coffman-Kundu-Wootters inequality is fulfilled in both models we studied, and  the obtained entanglement distribution indicates the presence of a relatively large fraction of long-range entanglement in the critical region for both Ising and XXZ models in both 1D and 2D.

Our work may be extended into several directions: In order to more deeply analyze the numerical possibilities of the translationally invariant MPS algorithm one needs to implement it efficiently for negative parameter values, that is negative $\kappa$ from Section~\ref{numer_methods}. Furthermore, it is desirable to have codes where the symmetries of the  ground states can be prescribed from the outset. Such work is under way.

Moreover, for more complete entanglement characterization of the models it is important to take into account other entanglement measures and characteristics, such as fidelity~\cite{PhysRevLett.103.170501}, global entanglement~\cite{PhysRevA.81.032304}, and entanglement spectrum~\cite{PhysRevLett.101.010504}. Another promising direction is the analysis of the complementarity of the entanglement~\cite{Jakob2010827} in many-body systems.
And, of course, it would be interesting to extend our studies to higher spins.

\appendix

\section*{Appendix: Entanglement measures}\label{entanglement}

In this appendix we briefly review well known definitions for various bipartite entanglement measures.

The first two are the one-site entanglement entropy $S_1$ and one-tangle $\tau_1$, which are obtained directly from the single-site reduced density matrix.
The \textit{entanglement entropy}~\cite{nielsen:book} for bipartite pure states $|\psi_{12}\rangle$ is the von Neumann entropy of the reduced density matrix
\begin{equation}
S(|\psi_{12}\rangle)=\mathcal{S}(\rho_1)=\mathcal{S}(\rho_2),
\end{equation}
with the reduced density matrices $\rho_1= {\rm Tr}_2 (\rho_{12})$ and \ $\rho_2= Tr_1 (\rho_{12})$; \ $\rho_{12}=|\psi_{12}\rangle\langle\psi_{12}|$ and Tr$_i$ indicates a trace over the subsystem $i$.
The von Neumann entropy $S$ of a density matrix $\rho$ is calculated from its eigenvalues~\cite{nielsen:book} $\lambda_i$:
\begin{equation}
S(\rho)=-\rho\log_2\rho=-\sum_i \lambda_i\log_2\lambda_i.
\end{equation}

In the main text we use $S_1=S(\rho_1)$. The one-tangle~\cite{PhysRevA.61.052306} is also calculated from one-site reduced density matrix:
\begin{equation}
\tau_1(\rho_1)=4\det \rho_1.
\end{equation}
The von Neumann entropy is connected to the one-tangle through the relation~\cite{RevModPhys.80.517}
\begin{equation}
\mathcal{S}(\rho_1)=h\left( \frac{1}{2} + \frac{\sqrt{1-\tau_1(\rho_1)}}{2} \right),
\end{equation}
where $h(x)=-x\log_2 x - (1-x)\log_2(1-x)$ denotes the binary entropy function.

Next we mention measures obtained from the two-site reduced density matrix $\rho_{12}$.
A simple measure of bipartite entanglement in a mixed state is the {\it entanglement of formation}, $E_F$~\cite{PhysRevA.54.3824}. It counts the minimum number of maximally entangled states (Bell states) needed to construct a given state using only local operations and classical communication (LOCC)~(for details see~\cite{PhysRevA.54.3824, nielsen:book}).
The entanglement of formation can be calculated from the concurrence of formation $C_F$~\cite{PhysRevLett.78.5022,
PhysRevLett.80.2245}:
\begin{equation}
E_F=h\left(\frac{1}{2}+\frac{\sqrt{1-C_F^2}}{2} \right),
\end{equation}
where $h(x)$ denotes the binary entropy function.
The \textit{concurrence} of formation ~\cite{ PhysRevLett.80.2245} is an entanglement measure for mixed states of two qubits, defined as
\begin{equation}
\label{conc}
C_F(\rho)=\max(0,\lambda_1-\lambda_2-\lambda_3-\lambda_4),
\end{equation}
where $\lambda_1,\lambda_2,\lambda_3,\lambda_4$ are the eigenvalues in decreasing order of the Hermitian matrix
\begin{equation}
R=\sqrt{\sqrt{\rho_{12}}\tilde{\rho}_{12}\sqrt{\rho_{12}}}
\end{equation}
with $\tilde{\rho}_{12}=(\sigma_y\otimes\sigma_y)\rho_{12}^*(\sigma_y\otimes\sigma_y)$. Here $\rho^*_{12}$ is the complex conjugate of the two-site density matrix $\rho_{12}$.
Alternatively, $\lambda_i$ are the square roots
of the singular values of the non-Hermitian matrix $\rho_{12}\tilde{\rho}_{12}$.
The concurrence is zero for a product state and one for a maximally entangled state.

Another type of concurrence, the concurrence of assistance $C_A$, was introduced in connection with the \textit{entanglement of assistance} $E_A$~\cite{DiVincenzo1999}. $C_A$ is obtained from~\cite{Laustsen2003}:
\begin{equation}
C_A=\lambda_1+\lambda_2+\lambda_3+\lambda_4.
\end{equation}
The entanglement of assistance measures the maximal bipartite entanglement which be obtained while doing measurements on the rest of the spins. The idea of entanglement of assistance originates from the analysis of tripartite systems, described by a state $|\psi^{123}\rangle$. By varying the measurement on party 3, the `helper' 3 is able to influence the mixed state of parties 1 and 2~\cite{PhysRevA.72.052317}. In order to use $E_A$ in practice one must be able to perform a maximization over all different measurement strategies, thus this measure is difficult to calculate. However, there exist easily calculable bounds on $E_A$: upper bounds on $E_A$ are the entropic bound, the fidelity bound, and concurrence bound $C_A$~\cite{DiVincenzo1999}. The latter is used in the present paper.

The \textit{localizable entanglement} $E_L$~\cite{PhysRevLett.92.027901} is defined as the maximal amount of entanglement that can be localized (on average) between two spins while doing only local measurements on the rest of the spins in the environment. $E_L$ cannot be obtained from the reduced density matrix alone, thus it is able to describe characteristics of the wave function that are not captured by two-point correlation functions, e.g. exotic phases like topological orders. The calculation of $E_L$ is not a trivial task since one needs to optimize over all possible local measurement strategies, nevertheless it is possible to obtain bounds on $E_L$ using only two-point correlation functions~\cite{PhysRevLett.92.027901}.

The upper bound for $E_L$ is the concurrence of assistance $C_A$, and the lower bound is obtained from the maximal two-point correlation function,
\begin{equation}
\max(|Q_{12}^{xx}|,|Q_{12}^{yy}|,|Q_{12}^{zz}|) \leq E_L \leq C_A,
\end{equation}
where $Q_{12}^{\alpha \beta}(|\psi\rangle\langle \psi|) = \langle \psi |\sigma_1^{\alpha}\otimes \sigma_2^{\beta} |\psi\rangle -
\langle \psi |\sigma_1^{\alpha}\otimes 1_2|\psi\rangle \langle \psi |1_1 \otimes\sigma_2^{\beta}|\psi\rangle$  and $\sigma_\alpha$ are the Pauli spin matrices.

The \textit{negativity}~\cite{PhysRevA.65.032314} is an `easy-to-compute' measure defined as
\begin{equation}
\mathcal{N}(\rho_{12})=\frac{|| \rho^{\Gamma_1}||_1-1}{2},
\end{equation}
where $\rho_{12}^{\Gamma_1}$ is the partially transposed density matrix $\rho_{12}$ with respect to subsystem $1$. And $||\rho_{12}||_1={\rm Tr}\sqrt{\rho_{12}^{\dagger}\rho_{12}}$ is the trace norm.
$||\rho_{12}||_1$ is calculated as a sum of the singular values of $\rho_{12}$.
A measure closely related to the negativity  is the \textit{logarithmic negativity}~\cite{PhysRevLett.95.090503},
\begin{equation}
E_N(\rho)=\log_2(||\rho^{\Gamma_1}||_1).
\end{equation}

A simple form of bipartite entanglement is the entanglement between two neighboring spins and the other spins of the system. This measure is called  \textit{local entanglement}~\cite{1367-2630-8-4-061}.
The two-site local entanglement $S_{\rm loc}$ is obtained by tracing out all spin degrees of freedom of the system except the two nearest-neighbour spins and then calculating the von Neumann entropy of the resulting reduced density matrix $\rho_{12}$,
\begin{equation}
S_{\rm loc}=S(\rho_{12}).
\end{equation}

Another entanglement measure, which can be used if we have available a tensor network representation of the state in conventional form, is the \textit{bipartite entanglement per bond} $S_{\rm PB}$~\cite{PhysRevA.81.032304}. It is obtained from the bond vectors~\cite{vidal:3} (see also section 2) connecting two neighboring sites.
The bond vectors contain essential entanglement information of the system.
The entanglement per bond $S_{\rm PB}$ is given by
\begin{equation}
S_{\rm PB}=\sum_i \lambda_i^2\log_2 \lambda_i^2.
\end{equation}
where the components of the bond vectors are normalized such that $\sum_i \lambda_i^2=1$.

There are other entanglement measures like fidelity~\cite{PhysRevLett.103.170501}, global entanglement~\cite{PhysRevA.81.032304}, entanglement spectrum~\cite{PhysRevLett.101.010504} with Schmidt gap, which also can be used to analyze entanglement in many-body systems. These measures are not considered in the present text.

\bibliography{refs,spins}

\end{document}